\documentclass[twocolumn,trackchanges]{aastex701}

\def\zem{$z_{\rm em}$~}

\def\mgii{Mg\,{\sc ii}}

\def\aliii{Al\,{\sc iii}}
\def\feiii{Fe\,{\sc iii}~}
\def\feii{Fe\,{\sc ii}}

\def\kms{km~s$^{-1}$}
\usepackage{csquotes}
\usepackage{lineno}
\usepackage{float}

\usepackage{array}      % To customize column width
\usepackage{lipsum} 
\begin{document}

\title{Recurrent Multi-year Mg II BAL Variability in SDSS J1333+0012}

\author[orcid=0000-0001-5937-331X,sname='Vivek']{M. Vivek}
%\altaffiliation{Kitt Peak National Observatory}
\affiliation{Indian Institute of Astrophysics, Koramangala II block, Bangalore 560034, India}
\affiliation{Astronomisches Rechen-Institut, Zentrum fur Astronomie der Universitat Heidelberg, Monchhofstr. 12-14, D-69120 Heidelberg,
Germany}
\email[show]{vivek.m@iiap.res.in}  

\author[orcid=0009-0001-2178-4022, sname='Aromal']{P. Aromal} 
\affiliation{Physics and Astronomy Department, University of Western Ontario, 1151 Richmond Street, London, N6A 3K7, Ontario, Canada }
\affiliation{Institute for Earth and Space Exploration, Western University, 1151 Richmond Street, London, ON N6A 3K7, Canada }
\email{apathaya@uwo.ca}

\author[orcid=0000-0002-9062-1921,sname='Srianand']{R. Srianand}
\affiliation{ IUCAA, Postbag 4, Ganeshkind, Pune 411007, India}
\email{anand@iucaa.in}

\author[orcid=0009-0008-6712-8734,sname='K.A. Anjali']{K.A. Anjali}
\affiliation{Indian Institute of Astrophysics, Koramangala II block, Bangalore 560034, India}
\email{anjali.ka@iiap.res.in}

\author[orcid=0000-0001-6217-8101,sname='S.C. Gallagher']{S. C. Gallagher}
\affiliation{Physics and Astronomy Department, University of Western Ontario, 1151 Richmond Street, London, N6A 3K7, Ontario, Canada }
\affiliation{Institute for Earth and Space Exploration, Western University, 1151 Richmond Street, London, ON N6A 3K7, Canada }
\email{sgalla4@uwo.ca}

\author[orcid=0000-0003-4592-447X, sname='Rachana']{Rachana}
\affiliation{Indian Institute of Astrophysics, Koramangala II block, Bangalore 560034, India}
\affiliation{Joint Astronomy Programme, Department of Physics, Indian Institute of Science, Bangalore 560012, India}
\email{rachana2022@iisc.ac.in}
% \author{River Europe}
% \affiliation{University of Heidelberg}
% \email{fakeemail4@google.com}

% \author[0000-0000-0000-0003,sname=Asia,gname=Mountain]{Asia Mountain}
% \altaffiliation{Astrosat Post-Doctoral Fellow}
% \affiliation{Tata Institute of Fundamental Research, Department of Astronomy}
% \email{fakeemail5@google.com}

% \author[0000-0000-0000-0004]{Coral Australia}
% \affiliation{James Cook University, Department of Physics}
% \email{fakeemail6@google.com}

% \author[gname=IceSheet]{Penguin Antarctica}
% \affiliation{Amundsen–Scott South Pole Station}
% \email{fakeemail7@google.com}

% \collaboration{all}{The Terra Mater collaboration}

%% Use the \collaboration command to identify collaborations. This command
%% takes an optional argument that is either a number or the word "all"
%% which tells the compiler how many of the authors above the command to
%% show. For example "\collaboration[all]{(DELVE Collaboration)}" wil include
%% all the authors above this command.
%%
%% Mark off the abstract in the ``abstract'' environment. 
\begin{abstract}

We present a long-term spectroscopic study of Broad Absorption Line (BAL) variability in the quasar J1333+0012 (\zem = 0.9197) using an extensive multi-epoch dataset spanning nearly two decades. The dataset comprises multi-epoch observations from multiple observatories, delivering nearly uniform temporal coverage of the \mgii\ BAL profile, with close to one spectroscopic observation per year spanning the period from 2008 to 2025. { We identify recurrent multi-year variability in the BAL absorption strength, characterized by a rest-frame timescale of $\sim$4 yr. Because the well-sampled baseline covers only a small number of candidate cycles, we present J1333+0012 as a candidate quasi-periodic BAL system.} In contrast, contemporaneous optical photometric light curves from CRTS, Pan-STARRS, ZTF, and PTF show no statistically significant correlated variability on comparable timescales. { However, ionization-driven variability cannot be ruled out, as the observed optical continuum does not directly trace the unobserved extreme-UV ionizing continuum.} The coherence of the variability across distinct velocity components, coupled with the absence of large-scale profile reshaping, is consistent with a geometric origin. We interpret the observed behavior as modulation of the line-of-sight absorption by the rotation of an inhomogeneous shielding-gas structure.

\end{abstract}

%% Keywords should appear after the \end{abstract} command. 
%% The AAS Journals now uses Unified Astronomy Thesaurus (UAT) concepts:
%% https://astrothesaurus.org
%% You will be asked to selected these concepts during the submission process
%% but this old "keyword" functionality is maintained in case authors want
%% to include these concepts in their preprints.
%%
%% You can use the \uat command to link your UAT concepts back its source.

%% From the front matter, we move on to the body of the paper.
%% Sections are demarcated by \section and \subsection, respectively.
%% Observe the use of the LaTeX \label
%% command after the \subsection to give a symbolic KEY to the
%% subsection for cross-referencing in a \ref command.
%% You can use LaTeX's \ref and \label commands to keep track of
%% cross-references to sections, equations, tables, and figures.
%% That way, if you change the order of any elements, LaTeX will
%% automatically renumber them.

\section{Introduction} 
Outflows from active galactic nuclei (AGN) are now recognized as a fundamental component of black hole–galaxy coevolution. AGN-driven winds carry mass, momentum, and energy from near the supermassive black hole into the host galaxy. These winds can influence star formation, redistribute interstellar gas, and explain the tight correlation between black hole mass and the properties of their host galaxies \citep{Fabian2012,Kormendy2013}. AGN outflows have been observed over a wide range of spatial scales and wavelengths, ranging from ultra-fast X-ray winds at sub-parsec scales to ionized and molecular outflows at kiloparsec scales, indicating a complex, multi-phase feedback cycle \citep{Tombesi2010, cicone2014}.

Broad Absorption Line (BAL) quasars, identified by blue-shifted broad absorption lines (exceeding several thousand \kms) in the optical and ultraviolet spectra of AGN provide the clearest signatures of powerful AGN outflows along our line of sight \citep{weynman91}. BAL quasars constitute 10–20\% of optically selected quasar samples, but this fraction may be higher due to selection effects, suggesting that BAL outflows are a common feature of quasar activity \citep{hewett2003, Hiremath2025}. Based on the ionic species present, BAL quasars are commonly classified into high-ionization BALs (HiBALs), which show only high-ionization lines, and low-ionization BALs (LoBALs), which additionally exhibit absorption from species such as \mgii\ and \aliii. A rare subset, FeLoBALs, further display absorption from excited and metastable \feii\ and \feiii\ transitions \citep{Hall2002}.

Time variability of BAL troughs has emerged as a powerful diagnostic of the physical origin, structure, and location of quasar outflows. Multi-epoch spectroscopic studies have revealed that BALs frequently vary in equivalent width, depth, and profile shape on rest-frame timescales ranging from months to years \citep{filiz13,vivek14, green2023, Aromal2025}. Two main types of mechanisms are often used to explain this variability : changes in the ionization state of the absorbing gas caused by continuum fluctuations, and, geometric effects such as transverse motion of inhomogeneous absorbing structures across the line of sight. Statistical studies have shown that variability often occurs coherently over limited velocity intervals and is more pronounced at higher outflow velocities \citep{filiz13, aromal2023}, favoring models involving clumpy or filamentary winds rather than smooth, homogeneous flows. Despite extensive monitoring campaigns, however, most BAL variability appears stochastic, and clear evidence for periodic or quasi-periodic behavior remains elusive.

SDSS J133356.02+001229.1 (hereafter, SDSS J1333+0012; \zem = 0.9197) is a LoBAL quasar famously recognized for its variability in the \mgii\ absorption line.
\citet{vivek12} first reported the transience of its \mgii\ BAL features, noting the complete disappearance of a low-velocity component (17,000 \kms) and the emergence of a new, high-velocity component at approximately 28,000 \kms\,
 that evolved significantly over just a few years. In a follow-up study, \citet{vivek2018} utilized follow-up observations  to reveal even more dramatic behavior, where the high-velocity component re-emerged and nearly disappeared again within timescales of just a few days to 4.2 years in the quasar rest frame. These studies attributed the absorption line variability to changing ionizing flux, potentially caused by variable shielding or individual gas clouds crossing the line of sight.

{ In this paper, we present a comprehensive analysis of nearly two decades of spectroscopic monitoring of J1333+0012. We combine moderate-resolution spectra obtained from several facilities to construct one of the most detailed long-term \mgii\ BAL variability datasets for a single quasar. By focusing on the \mgii\ BAL profile, we examine how the absorption strength and velocity structure evolve over time across different trough components. Our analysis reveals recurrent multi-year BAL equivalent-width variations with a characteristic rest-frame timescale of approximately four years. Because the well-sampled baseline covers only a small number of candidate cycles, we treat this behavior as candidate quasi-periodic variability rather than as a confirmed periodic signal.
}
%In this paper, we present a comprehensive analysis of nearly two decades of spectroscopic monitoring of J1333+0012. We combine observations from various facilities to create one of the most detailed long-term BAL datasets for a single quasar. By focusing on the \mgii\ BAL profile, we examine how the absorption strength, and velocity structure  change over time across different trough components. Our analysis shows distinct quasi-periodic patterns in the BAL variability, providing new insights into the structure and dynamics of the outflowing gas.

\section{Data \& Analysis} 
Our spectroscopic observations significantly extend the temporal baseline presented in \citet{vivek12,vivek2018}. The full dataset includes  24 spectra from the Sloan Digital Sky Survey (SDSS), the IUCAA Girawali Observatory (IGO), the Southern African Large Telescope (SALT), and the Dark Energy Spectroscopic Instrument (DESI). We refer  to the earlier papers for details of the observations and data reduction procedures for spectra obtained up to 2016. Following 2016, we resumed monitoring in 2018 and continued obtaining spectra primarily with SALT using the same instrumental configuration as in our previous campaigns. 
The SALT observations cover the wavelength range 4500$-$7500 \AA\ with a spectral resolution of approximately 300 \kms, and we secured at least one epoch per year to ensure consistent temporal sampling of the \mgii\ BAL region. For all these data, we used data reduction steps as explained in \citet{aromal2023}. In addition to the SALT monitoring, one further epoch each was obtained from SDSS and DESI, providing independent, homogeneously reduced survey-quality spectra. Details of the  observing log, including dates, exposure times, and instrumental setups, are provided in the Appendix Table.~\ref{tab_obs_log}.

For continuum normalization, we first fit a continuum to the highest-SNR SDSS spectrum (obtained on MJD : 51955) using PyQSOFit \citep{Wu2022}, which performs a physically motivated spectral decomposition including a power-law continuum, broad emission-line components, and blended pseudo-continuum features such as Fe II emission. For all remaining epochs, we did not refit the continuum independently. Instead, each epoch was aligned to the reference continuum by allowing two controlled transformations: (1) an overall multiplicative scaling factor and (2) a linear spectral tilt while pivoting around a chosen wavelength. The scaling accounts for differences in absolute flux calibration, slit losses, variable observing conditions, and intrinsic continuum variability. The tilt corrects for small differences in spectral slope that can arise from instrumental response variations or residual calibration mismatches. During the alignment process, care was taken to ensure that the continuum regions on both sides of the \mgii\ BAL trough were consistently matched to those of the reference spectrum. Once all epoch spectra were aligned to the reference spectrum using the transformation above, the reference continuum model was divided by each transformed spectrum to produce the normalized spectra for subsequent analysis. This normalization scheme ensures a uniform continuum definition across all epochs and substantially reduces systematic errors introduced by independent continuum fits. The continuum fits for each epoch and the corresponding continuum-normalized spectra are presented in the Appendix Fig.~\ref{fig_continuum_plots}.

Equivalent widths (EWs) of the \mgii\ BAL components were measured from the continuum-normalized, rest-frame aligned spectra. We adopted the same definition of absorption components as in \citet{vivek2018}, dividing the profile into five regions :  R (5028 \AA $-$ 5100 \AA; v $\approx$ -15200 to -19237 \kms ), B1(4925 \AA $-$ 5000 \AA; v $\approx$ -20800 to -24980 \kms) , B2 (4840 \AA $-$ 4925 \AA;  v $\approx$ -24980 to -29730 \kms), and B3(4745 \AA $-$ 4840 \AA; v $\approx$ -29730 to -35030 \kms)  components, as well as the combined B-total region.
%\Aromal{Also, mention the corresponding velocities?}

\section{Results}
\begin{figure}
    \centering
    \includegraphics[width=1.0\linewidth]{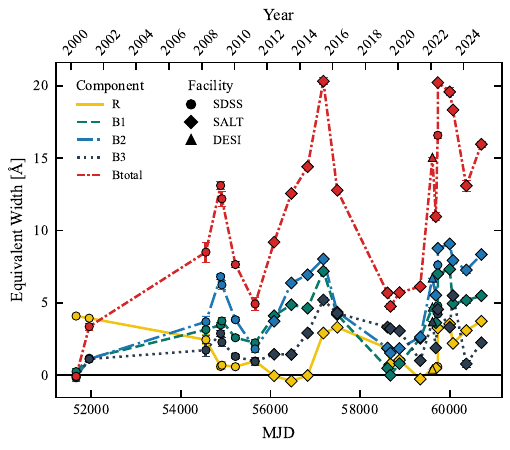}
    \caption{The equivalent widths of the R (solid/yellow), B1 (dashed/green), B2 (dash–dot/blue), B3 (dotted/dark blue), and B-total (dash-dot-dot/red) absorption components are shown as a function of MJD. Marker shapes denote the observing facility: SDSS (circle), IGO (square), SALT (diamond), and DESI (triangle up).}
    \label{fig_EW_MJD}
\end{figure}
Fig.~\ref{fig_EW_MJD} shows the variation of the \mgii\ BAL equivalent widths as a function of observation date (MJD) for the individual absorption components R (solid/yellow), B1 (dashed/green), B2 (dash–dot/blue), B3 (dotted/dark blue), and B-total (dash-dot-dot/red) region. All components exhibit significant long-term variability over the full monitoring period, with the B-total component displaying the largest amplitude changes.

The R component displays comparatively modest but structured variability. It begins at EW$\approx4$\AA\ during 2001$-$2003, declines substantially over 2008$-$2013, and then re-strengthens beginning in 2015. This is followed by another weakening phase around 2020$-$2021 and a subsequent recovery in the most recent epochs.  In contrast, the B absorption components exhibit substantially larger amplitude variations. Both B1 and B2 rise dramatically from near-zero levels in 2001 and undergo alternating strengthening and weakening phases, with pronounced maxima around 2009, 2015, and 2023, and relative minima near 2011 and 2019. The B3 component generally tracks the overall behavior of the blue complex, albeit with somewhat lower amplitude. A caveat is that, for the SALT spectra obtained between 2019 and 2022, the wavelength coverage does not fully encompass the B3 component, and therefore its equivalent width measurements during these epochs should be interpreted with caution. %The B-total equivalent width, which integrates the entire blue absorption complex, most clearly illustrates the long-term evolution of the outflow strength. It reveals at least three distinct cycles of strengthening and weakening over the monitoring period of 12.9 years in the quasar rest-frame. The recurrence of these multi-year variations strongly favors a cyclic or quasi-periodic variation of the outflow rather than purely stochastic fluctuations. 
{ The B-total equivalent width, which integrates the entire blue absorption complex, most clearly illustrates the long-term evolution of the outflow strength. It shows recurrent phases of strengthening and weakening over the 12.9 yr rest-frame monitoring baseline. The strongest maxima occur around 2009, 2015, and 2023, with relative minima near 2011 and 2019. This pattern is suggestive of a characteristic multi-year recurrence timescale.}% However, because only a small number of candidate cycles are covered, the recurrence should not by itself be interpreted as a secure periodic signal. We therefore use the term candidate quasi-periodic variability throughout this work.}

%\Aromal{Shall we label the top axis of Fig 1 with years as it will be easier to follow with the variability description?} 
%\Anand{The top axis is crowded. You may print the label every two years.}

We then performed a periodicity analysis on the measured equivalent widths of the B-total component. Beginning in 2008, the source has been monitored with a nearly annual cadence, providing substantially more uniform temporal coverage compared to the earlier two SDSS epochs. To minimize biases arising from sparse and uneven sampling, we restrict the periodicity analysis to the EW measurements obtained from 2008 onward. %A Lomb–Scargle periodogram \citep{scargle1982} reveals a prominent peak at a period of 1586 days, with a false-alarm probability (FAP) of 5.5$\times$10$^{-4}$, indicating that the detected signal is statistically significant.
{ A Lomb–Scargle periodogram reveals a prominent peak at a period of 1586 days, with a  white-noise false-alarm probability (FAP) of $5.5\times10^{-4}$. However, this  FAP should be interpreted cautiously because stochastic red-noise processes, such as DRW variability, can produce apparently significant periodogram peaks in sparsely sampled light curves.}

\begin{figure*}
    \centering
    \includegraphics[width=1\linewidth,trim=36 0 45 0, clip]{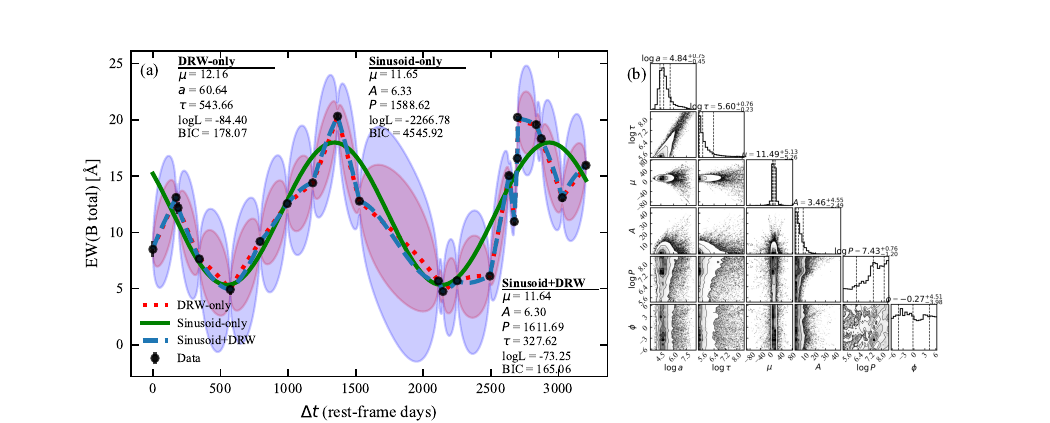}
    \caption{a) Rest-frame evolution of the B-total equivalent width for J1333+0012 (MJD $>$ 54000), compared with three competing variability models: a damped random walk (DRW; red), a purely sinusoidal model (green), and a combined sinusoid + DRW  model (blue). Shaded regions indicate the 1$\sigma$ predictive uncertainty of the Gaussian Process components. The DRW model assumes an exponential covariance kernel with characteristic timescale $\tau$, while the sinusoidal component is parameterized by amplitude A and period  P. The inset summaries list the maximum-likelihood parameters and corresponding Bayesian Information Criterion (BIC) values for each model. (b) Corner plot showing the posterior distributions of the six parameters of the sinusoid + DRW model obtained from MCMC sampling. }
    \label{fig:periodicity}
\end{figure*} 
To investigate the nature of the variability in the B-total equivalent width, we compared three competing models: (1) a purely stochastic damped random walk (DRW) model, (2) a purely deterministic sinusoidal model with white noise, and (3) a hybrid model combining a sinusoidal component with a DRW model. 

We first fitted the B-total EW time series with two baseline models: a purely stochastic damped random walk (DRW) model and a purely deterministic sinusoidal model. The DRW fit was implemented as a Gaussian Process (GP), which provides a likelihood-based framework for modeling correlated  variability with full propagation of measurement uncertainties. We used the celerite2 implementation of GPs \citep{gpr2006}, adopting an exponential covariance kernel (RealTerm) appropriate for a DRW process, k$\left(\Delta t \right) $= a e$^{-|\Delta t|/\tau}$ , where `a' denotes the DRW process variance and $\tau$ is the characteristic damping timescale.  Model parameters  were optimized by maximizing the GP log-likelihood  with physically motivated bounds on $\tau$ (i.e [200,5000] days).
In parallel, we fitted a sinusoid-only model assuming independent Gaussian measurement errors, m(t)=$\mu$+Asin(2$\pi$t/P+$\phi$), where $\mu$ is the mean level, A the amplitude, P the period, and $\phi$ the phase.  Finally, we fitted a combined sinusoid+DRW model in which the sinusoidal function represents the mean trend, while the residual variability is modeled as a DRW process. This hybrid framework allows us to assess whether the observed variability is better described by purely stochastic red-noise behavior, coherent periodic modulation, or a combination of both.

Figure~\ref{fig:periodicity} shows the rest-frame temporal evolution of the B-total component EW  compared with three variability models. The sinusoid-only model captures coherent periodic behavior but does not account for correlated stochastic fluctuations. The shaded regions represent the 1$\sigma$  uncertainties of the Gaussian Process components. The inset parameter summaries and corresponding Bayesian Information Criterion (BIC) values indicate that the hybrid sinusoid+DRW model is statistically preferred over both the purely sinusoidal and DRW-only models (See also Fig.~\ref{fig:periodicity_B2}, which shows the same analysis for the B2 component alone, which does not suffer from any wavelength coverage issues.).
The variability is  best described as a quasi-periodic modulation superposed on stochastic red-noise variability. { The DRW-only model reproduces much of the smooth long-term evolution. For the B-total EW curve, the DRW-only model gives ${\rm BIC}=178.07$, while the sinusoid+DRW model gives ${\rm BIC}=165.06$, corresponding to $\Delta{\rm BIC}=13.01$ in favor of the hybrid model. Thus, the sinusoid+DRW model provides a better phenomenological description of the B-total EW variations.}  However, this improvement should not be interpreted as definitive evidence for a true periodic driver. Additional degrees of freedom can naturally improve the fit to stochastic light curves, and the available temporal baseline spans only a limited number of candidate cycles. We therefore regard the sinusoid+DRW fit as supporting candidate quasi-periodic behavior. In the DRW-only case, a timescale of  $\tau \sim$500 days is physically unlikely if the variability is driven purely by photoionization.  Studies of AGN continuum variability using optical light curves show that quasar DRW timescales are typically tens to hundreds of days in the rest frame \citep{macleod2010}. Since $\tau$ is often associated with the thermal timescale of the accretion disk, such a large value of $\tau$ $\sim$ 500 days, would imply an unusually massive supermassive black hole for this source. A more plausible interpretation is that the variability is linked to dynamical processes in the disk wind. Radiatively driven disk-wind simulations by \citet{proga2000} show that the outflow is inherently unstable and develops dense clumps (“density knots”) that form through line-driven instabilities and propagate outward. The formation and evolution of these structures occur on year-scale dynamical timescales, which could naturally explain the large effective $\tau$ inferred from the DRW modeling. { Additionally, long BAL EW damping timescale can also arise if the absorbing gas responds to continuum variations over a finite recombination or ionization-equilibration timescale \citep{Krolik1995,He2019}. In this case, the BAL EW curve would represent a smoothed or low-pass-filtered version of the driving ionizing continuum, and stochastic continuum variability could appear more coherent in the absorption response.}

\begin{figure}
    \centering
    \includegraphics[width=1\linewidth]{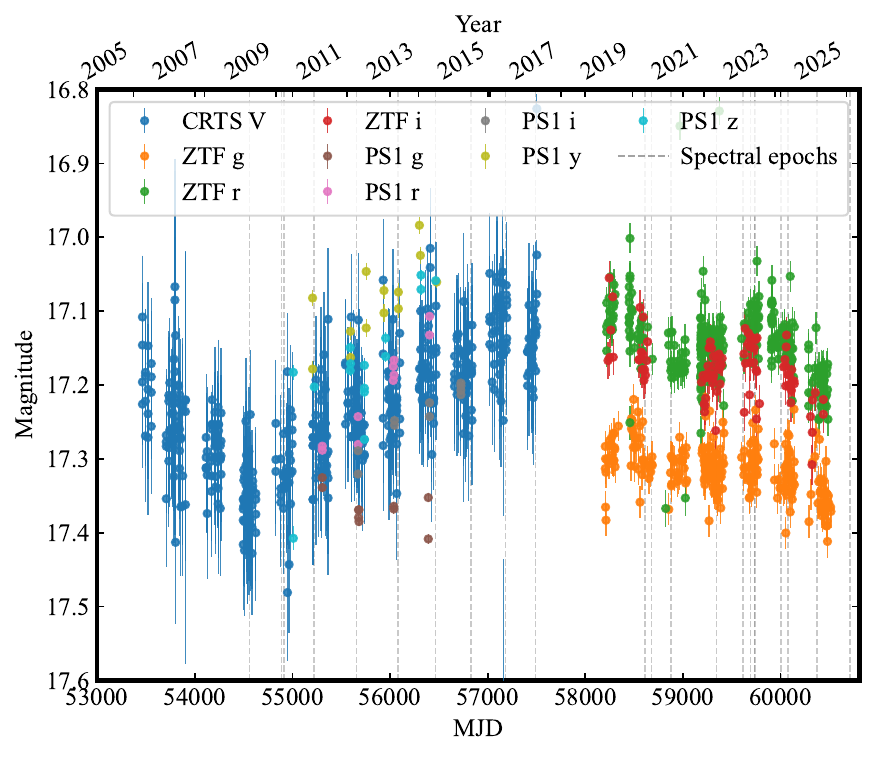}
    \caption{Optical light curve of SDSS J1333+0012 combining CRTS (V band), ZTF (g, r, i), and Pan-STARRS (g, r, i, y, z) photometry. Magnitudes are plotted in the observed frame as a function of MJD. Vertical dashed lines mark the spectroscopic epochs used for BAL equivalent width measurements.}
    \label{fig_lightcurve}
\end{figure}

The right panel presents the posterior distributions of the six parameters of the sinusoid+DRW model. The period is constrained to multi-year timescales, while the DRW damping timescale is substantially shorter, demonstrating a separation between coherent modulation and stochastic variability. The marginalized posterior distribution of the period
P is broad and skewed, indicating that the period is not tightly constrained by the data. While the posterior favors multi-year periods (of order $\sim$ 1500–2000 rest-frame days), it does not collapse to a narrow peak. This behavior is expected given the limited temporal baseline and the small number of observed cycles. For the same reason, the posterior distribution of the phase parameter, $\phi$ is also poorly constrained, reflecting the degeneracy between period and phase when the variability is sampled over a limited time span.  The strongest posterior correlation is observed between the DRW hyperparameters, a and 
$\tau$, reflecting the well-known variance–timescale degeneracy of stochastic processes. The sinusoidal amplitude,
A shows only weak correlations with the DRW parameters, indicating that the periodic component is not strongly absorbed by adjustments in the stochastic variability. %Together, these results indicate that while the data favor multi-year quasi-periodic modulation, the precise period and phase cannot yet be determined with high precision.
{ Together, these results indicate that the data are consistent with a multi-year recurrent component, but the period and phase are not tightly constrained by the present baseline. The posterior structure therefore supports a candidate quasi-periodic interpretation rather than a confirmed periodic detection.}
We compiled the long-term optical light curve of SDSS J1333+0012 using CRTS (V band), ZTF (g, r, i), and Pan-STARRS (g, r, i, y, z) photometry, spanning nearly two decades in the observed frame, as shown in Fig.~\ref{fig_lightcurve}. The source exhibits clear multi-year variability with a peak-to-peak amplitude of $\approx$0.3–0.6 mag. Spectroscopic epochs used for BAL equivalent width measurements are overplotted as vertical dashed lines. The continuum exhibits smooth secular evolution on multi-year timescales, broadly consistent with accretion-disk–driven variability. No clear correlation between continuum and BAL variability is evident. During the 2020–2025 interval, the BAL absorption shows strong variability, while the contemporaneous ZTF light curve displays only modest flux changes. This already suggests that the BAL evolution is unlikely to be driven solely by continuum ionization changes. Indeed, the ZTF g-band light curve is well described by a DRW process with $\tau$=36 days (rest-frame) and an asymptotic variability amplitude of SF$_{\infty}$=0.035 mag, values that are entirely typical of optical quasar variability in luminous quasars with L$_{bol}$ $\sim$ 10$^{45}$ $-$ 10$^{47}$ ergs s$^{-1}$\citep{macleod2010}. The inferred stochastic continuum variability amplitude is therefore only $\sim$3\%. By contrast, the BAL absorption strength exhibits much larger variations and quasi-periodic behavior on multi-year timescales, pointing to the presence of an additional mechanism beyond standard stochastic continuum fluctuations. %\Aromal{Does it help mentioning, say, the correlation coefficient of a fractional change in EW vs that of magnitude here?}
%\Aromal{Just a suggestion/idea : Is it good to do a DRW fit for the light curve here and compare with that of the value we get for BAL? What if this explains the DRW part of BAL variation which would probably suggest a separate mechanism for periodicity?}

{ Another possible way to test an ionization-driven scenario is to compare the \mgii\  BAL equivalent width directly with the continuum flux measured from the spectra. Such a test is most straightforward when the absorption is not strongly saturated, since in saturated BAL troughs the equivalent width may be governed primarily by covering fraction rather than by changes in ionic column density. Therefore, any EW–continuum comparison should be interpreted with this limitation in mind. In the present case, an additional practical limitation is that a significant fraction of our monitoring data were obtained with SALT, whose moving-pupil design and variable effective aperture prevent reliable absolute spectrophotometric calibration between epochs. Continuum fluxes measured directly from the SALT spectra are therefore affected by aperture-, throughput-, and observing-condition-dependent uncertainties that are unrelated to intrinsic quasar variability. For this reason, we do not consider the spectroscopic continuum fluxes to be a robust tracer of the intrinsic source continuum level for inter-epoch correlation analysis, and we avoid using them to compute the BAL EW–spectroscopic-continuum correlation.

As an alternative, we used the ZTF light curves as an external photometric continuum proxy for the post-2018 spectroscopic epochs. For the 11 epochs with contemporaneous ZTF coverage, we matched each spectroscopic MJD to the ZTF photometry and computed the median magnitude within a $\pm 50$ day window. We then compared these matched ZTF magnitudes with the \mgii\ BAL equivalent widths. This test does not show a statistically significant correlation ({ $\rho$ = -0.02 with p-value of 0.9}) between the optical continuum  and BAL strength. However, the number of matched epochs is still small, the scatter in the matched ZTF magnitudes is comparable to the typical photometric uncertainties, and the BAL equivalent width may not linearly trace ionic column density if the absorption is saturated or partially saturated. The available optical data do not show evidence for a simple one-to-one relation between observed optical continuum variability and \mgii\  BAL EW, but ionization variability driven by the unobserved EUV continuum, cannot be ruled out.}

\section{Discussion \& Conclusions} 
%Our long-term spectroscopic monitoring of SDSS J1333+0012 reveals a quasi-periodic variation in the BAL equivalent widths on a rest-frame timescale of approximately four years.
{ Our long-term spectroscopic monitoring of SDSS J1333+0012 reveals recurrent multi-year variability in the Mg II BAL equivalent widths, with a characteristic rest-frame timescale of approximately four years. The recurrence is visible most clearly in the B-total absorption complex and is also reflected, to varying degrees, in the individual velocity components. However, because the well-sampled baseline covers only a small number of candidate cycles, we do not claim a confirmed periodic signal. Instead, we identify J1333+0012 as a candidate quasi-periodic BAL system whose recurrence must be tested with continued monitoring.} The variability pattern is  recurrent across multiple cycles, yet it exhibits no clear correlation with the optical continuum light curve. Optical continuum does not directly trace the extreme-UV photons that govern the ionization balance of \mgii. Variability in the unobserved EUV continuum could thus drive changes in the absorber’s ionization state even in the absence of strong optical continuum variability. Accordingly, ionization-driven variability cannot be excluded based solely on the available optical monitoring data. 

An additional striking feature is the opposite behavior of the red and blue BAL components during the early epochs (2001–2009), where strengthening of one component coincides with weakening of the other. If we take at face value that variability in the unobserved EUV continuum drives the ionization changes, the contrasting behavior of the red and blue components can be interpreted as arising from their location on different sides of the column density–ionization parameter (log U) curve. Between 2001 and 2003, the red component is strong—consistent with higher density and lower ionization parameter—while the blue component is weak, likely corresponding to lower density and higher U, leading to opposite responses to the same continuum variations. From 2003 to 2008, as the continuum flux declines, the red component weakens or disappears while the blue strengthens. During 2009–2012, a substantial continuum brightening likely overionizes the red component, causing it to vanish; at the same time,  red component may provide partial shielding that prevents complete overionization of the blue component. Between 2012 and 2015, the red component remains weak initially before re-emerging as the continuum declines, accompanied by strengthening of the blue absorption. From 2015 to 2020, both components decrease despite increasing continuum flux, suggesting that simple overionization is no longer important. During 2020–2022, a continuum decline coincides with strengthening of both components, while the 2022–2025 interval shows relatively stable continuum and BAL properties. %However, this scenario demands substantial fine-tuning and an unlikely periodic variation of the EUV continuum, making ionization effects improbable as the primary driver of the periodic BAL evolution.
{ However, this scenario requires a structured absorber in which different velocity components respond differently to the same ionizing continuum. It may also require variations in the unobserved EUV continuum or in the shielding gas that are not directly traced by the optical light curve. We therefore cannot rule out ionization-driven variability, although a simple one-zone response to the observed optical continuum appears insufficient.}
%Additionally, it is well established that higher-velocity BAL components originate at smaller radii, consistent with their higher ionization parameters. The above picture assumes the opposite ordering, further highlighting the limitations of a purely ionization-driven interpretation.

An alternative explanation for the observed recurrent BAL variability is azimuthal structure within a rotating disk wind. If we interpret the candidate rest-frame recurrence timescale of 4.34 years as an orbital timescale, the implied radial location of the absorber can be estimated from Keplerian motion. Using the single-epoch virial black hole mass derived from the DESI spectra via the \citet{Shen2024} prescriptions, we obtain $M_{\rm BH} \sim 6.4 \times 10^{9},M_\odot$ (from \mgii) and $\sim 9.2 \times 10^{8},M_\odot$ (from H$\beta$). For a $\sim10^{9},M_\odot$ black hole, a 4.34-year orbital period corresponds to a radius of order $\sim15$ light-days, placing the absorber well inside the canonical \mgii\ broad-line region (BLR), whose reverberation lags are typically 50–100 light-days \citep{Homayani2020}. This would imply that the absorbing gas lies interior to, or deeply embedded within, the BLR, which is difficult to reconcile with standard BLR stratification and ionization structure. Therefore, unless the black hole mass is substantially larger than inferred, a simple picture of a single co-rotating cloud at the measured period is unlikely. A more plausible scenario may involve multiple azimuthal density enhancements within a structured disk wind, such that the line of sight intersects successive over-densities during rotation. However, maintaining strict periodicity over multiple cycles would still require long-lived, stable azimuthal structures, posing additional challenges for purely dynamical explanations %\Aromal{How about density knots formed by instabilities within a few year timescales as mentioned in Proga 2000, 2004?} {\color{red} MV : I have added the density-knot instability as a viable mechanism to explain the high DRW timescale of BAL variability}.

A more plausible explanation is variability driven by changes in a structured shielding gas located interior to the BAL outflow. If this shielding material orbits on a timescale comparable to the { observed $\sim$4.3-year recurrence timescale}, azimuthal variations in its column density could periodically modulate the EUV flux reaching the absorber. In this scenario, BAL variability is governed by geometric filtering of the ionizing continuum rather than by global continuum changes, and therefore a correlation with the observed optical flux is not expected. The early anti-correlated behavior of the red and blue components may reflect different velocity structures experiencing different shielding conditions, while the later correlated evolution suggests that both components eventually intercept similar ionizing flux. Thus, orbital modulation of a structured shielding layer provides a natural explanation for the { recurrent BAL variability} without requiring periodic intrinsic EUV luminosity changes. This interpretation is consistent with the shielding gas framework of \citet{vivek2018}, in which modest changes in column density and covering fraction can significantly alter the ionization state of low-ionization species such as \mgii, even without large intrinsic continuum variations. % \Anand{Since we have done some model of the shielding gas in Paper 2, you may refer the readers to that...}

In the cloud-crossing scenario, variability is attributed to individual absorbing clouds transiting the line of sight. However, this interpretation requires a mechanism that repeatedly launches clouds at specific velocities in a quasi-periodic manner, which appears physically unlikely.

As a speculative possibility, { recurrent BAL variability} could also arise from dynamical perturbations such as a binary black hole or nuclear star cluster inducing tidal structure in the wind, where the observed period may represent a higher harmonic of the orbital timescale (e.g., from a two-armed spiral), although this remains unconstrained by the present data. { Thus, the principal result of this work is the identification of the \mgii\ BAL in J1333+0012 as a potential candidate for exhibiting quasi-periodic \mgii\ absorption-line variability. Since periodic variability in BALs has not been observed to date, it is imperative to continue spectroscopic monitoring of J1333+0012 over several additional cycles to confirm and characterize this behavior. The persistence of the predicted strengthening and weakening pattern would support a quasi-periodic interpretation,  whereas significant deviations would favor stochastic variability.}%Further monitoring is essential to test the persistence of the periodicity and distinguish between geometric and dynamical origins.
%A spiraling outflow could in principle produce repeated BAL crossings if the line-of-sight velocity includes both rotational (v$_{\phi$}) and radial (vr) components; however, the near-constant recurrence period implies v$_r$≪v$_{\phi$, i.e., a tightly wound spiral. The observed 0.5–1 yr (rest-frame) duration of maximum absorption then implies a cloud size of $\sim$0.008–0.01 pc, comparable to the inferred spiral radius, making it difficult to simultaneously maintain compact, periodic crossings and extended absorption phases. Moreover, reproducing the early anti-correlated and later correlated behavior of the red and blue components within a single coherent spiral geometry is challenging. Together, these constraints render a simple spiral-crossing scenario less plausible without invoking finely tuned, long-lived azimuthal substructures.
\begin{acknowledgements}
The authors thank the anonymous referee for the constructive comments that have helped to improve the manuscript.    MV acknowledges support from DST-SERB in the form of a core research grant (CRG/2022/007884). P.A. and S.C.G. acknowledge support from the Western Research Chair program and Discovery Grant RGPIN-2021-04157 from the Natural Sciences and Engineering Research Council of Canada.
\end{acknowledgements}

\begin{appendix}

\begin{table*}[h]
\centering

\begin{tabular}{lccccc}
\hline
Epoch & Year/ & MJD & Exposure & Wavelength & Resolution \\
&Semester&&Time (s)&Coverage (\AA) & (km\,s$^{-1}$) \\

\hline
%SALT & 2018A & 58243 & 1200 & 4750--7790 & 300 \\
SALT & 2018B & 58613 & 1200 & 4750--7790 & 300 \\
SALT & 2019A & 58677 & 1300 & 4750--7790 & 300 \\
SALT & 2019B & 58879 & 1500 & 4750--7790 & 300 \\
SALT & 2021A & 59344 & 1500 & 4750--7790 & 300 \\
SALT & 2021B & 59691 & 1200 & 4750--7790 & 300 \\
SALT & 2022A & 59736 & 2100 & 4470--7510 & 300 \\
SALT & 2022B & 60001 & 1400 & 4470--7510 & 300 \\
SALT & 2023A & 60074 & 1600 & 4750--7790 & 300 \\
SALT & 2023B & 60372 & 1500 & 4470--7510 & 300 \\
SALT & 2024B & 60708 & 1400 & 4470--7510 & 300 \\
%SALT & 2025A & 60828 & 1500 & 4750--7790 & 300 \\
\hline
\end{tabular}

\caption{Log of latest SALT spectroscopic observations of SDSS J1333+0012. }
\label{tab_obs_log}

\end{table*}
\pagebreak

    \begin{figure*}
        \begin{tabular}{ll}
           \includegraphics[width=0.45\linewidth]{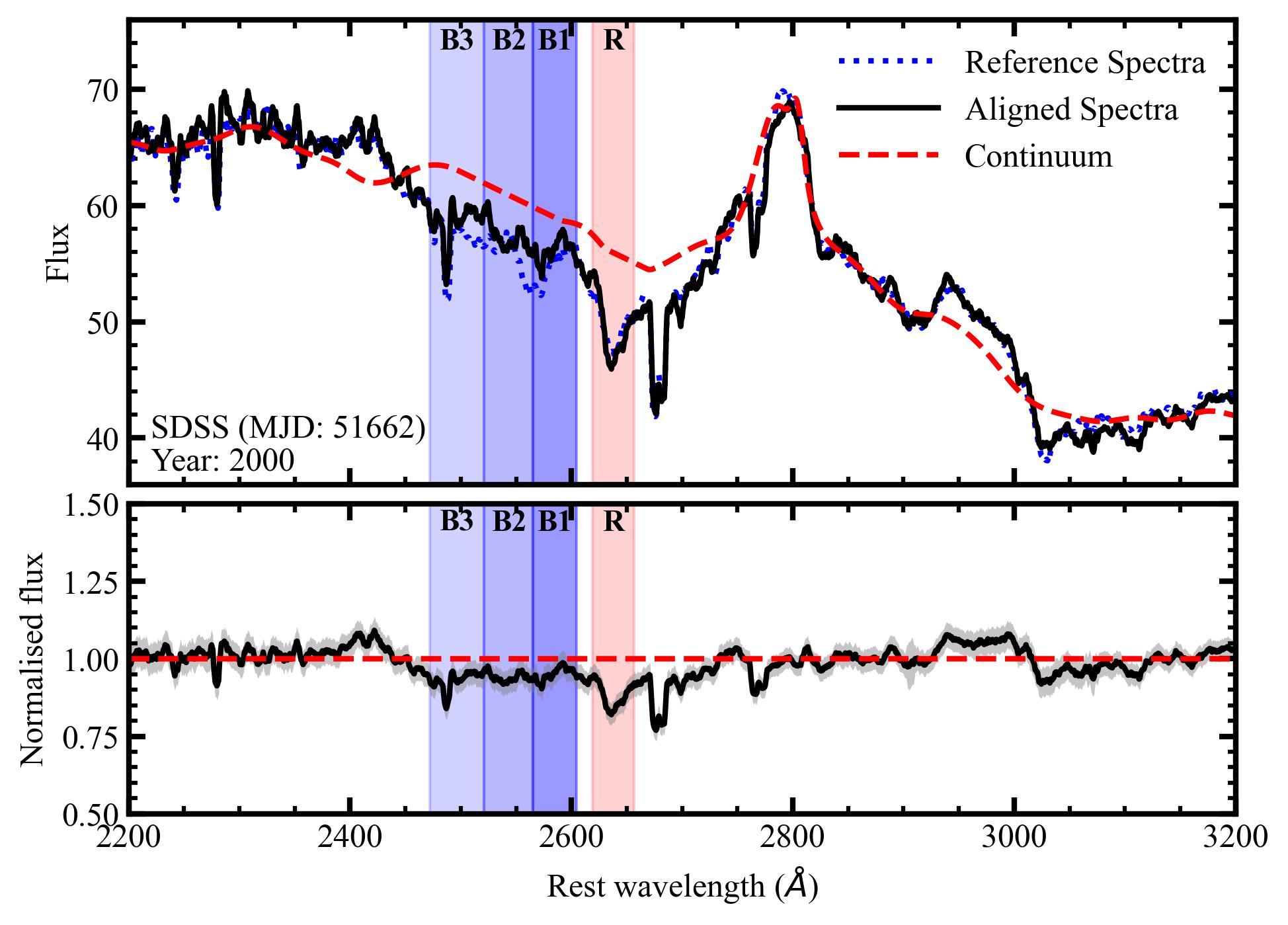}  & \includegraphics[width=0.45\linewidth]{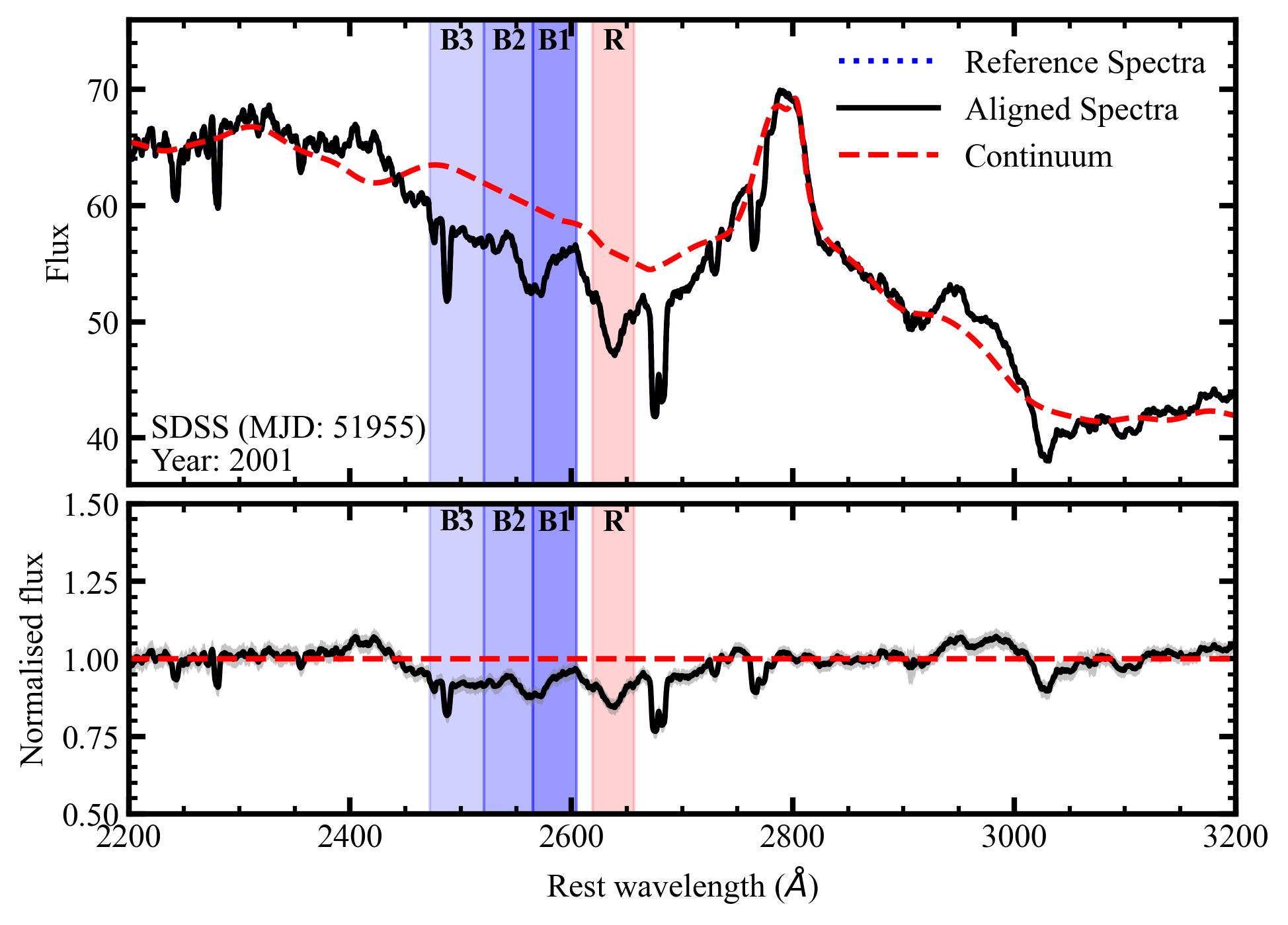} \\
             \includegraphics[width=0.45\linewidth]{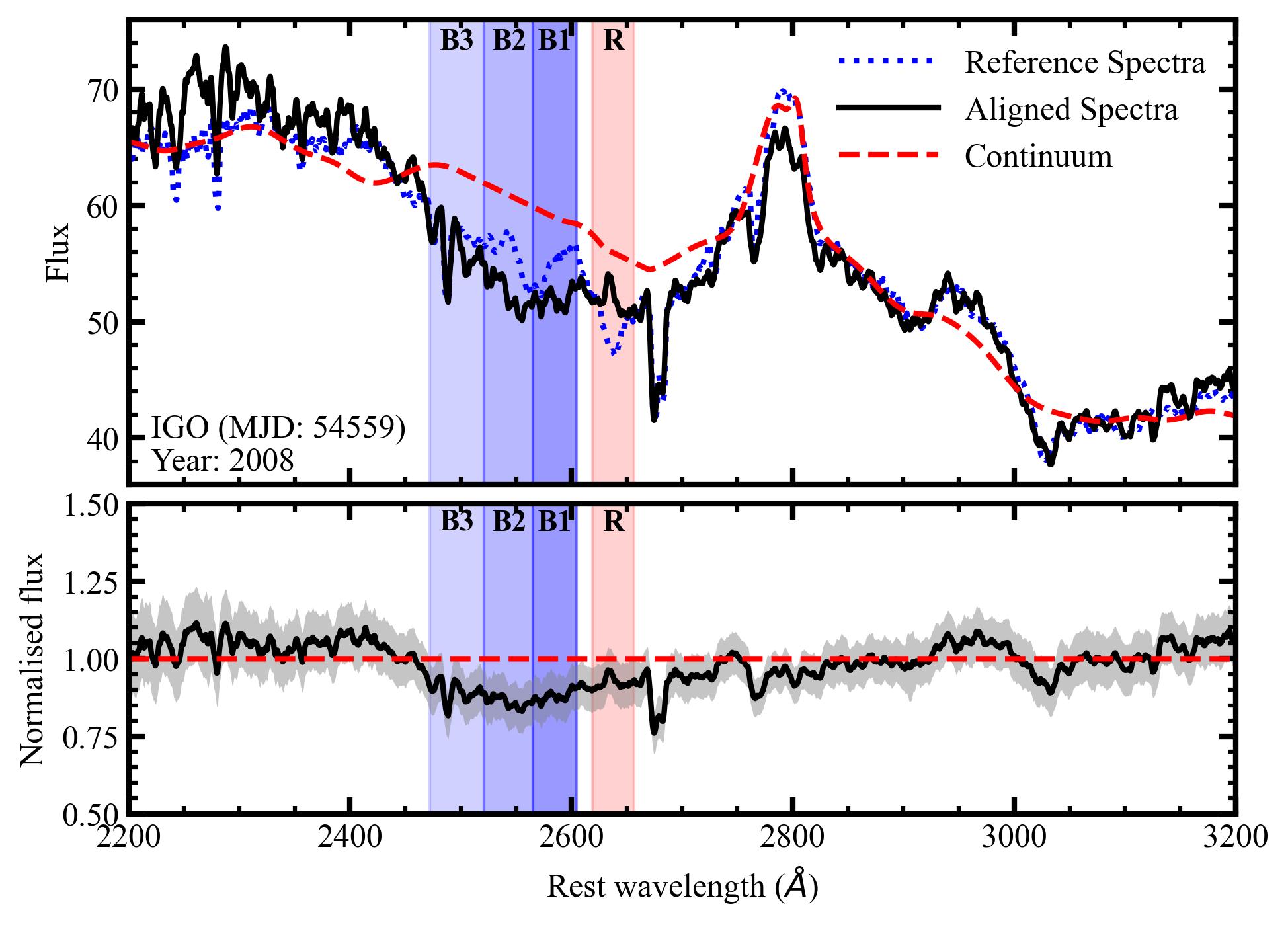}  & \includegraphics[width=0.45\linewidth]{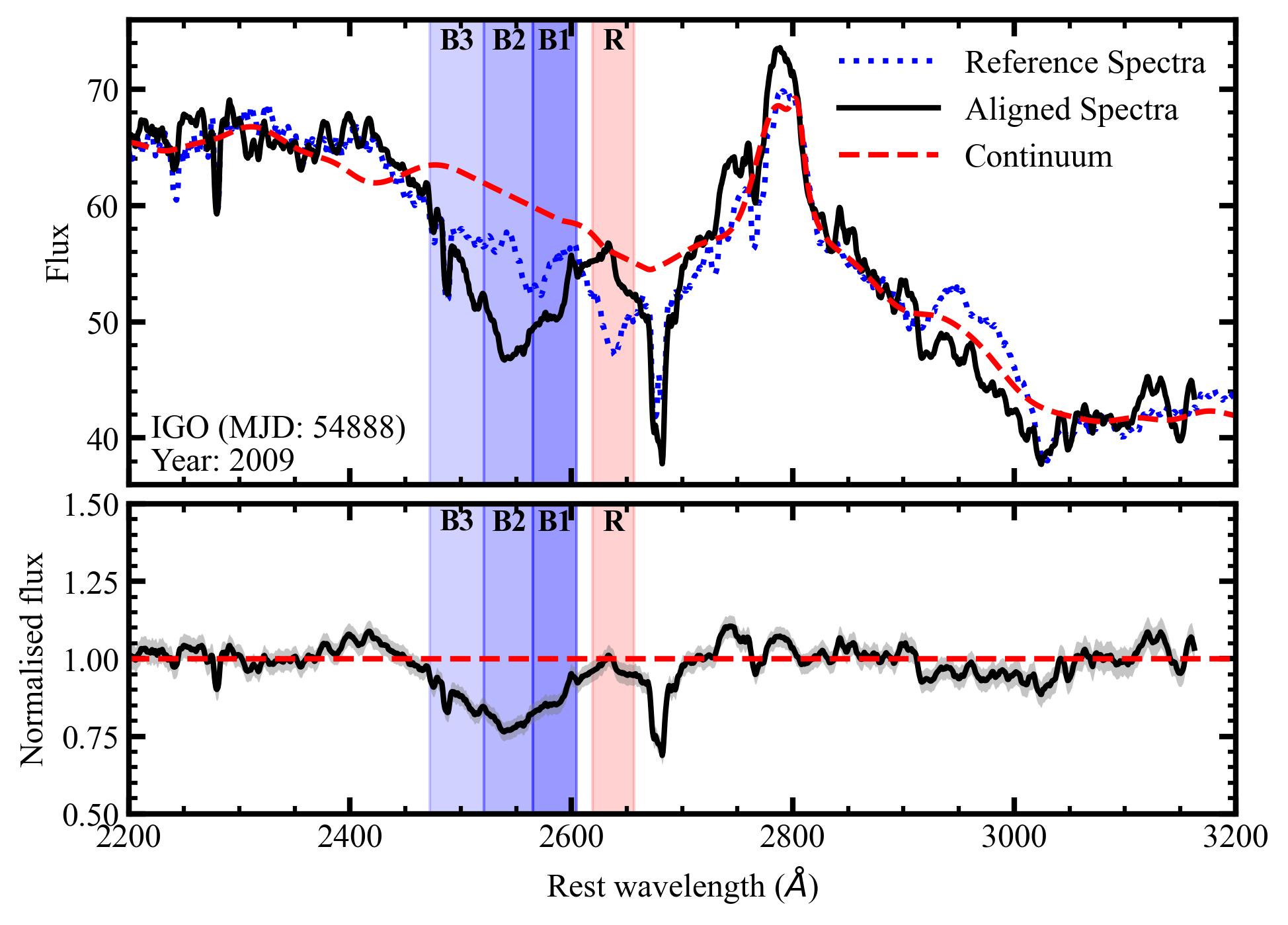} \\
             \includegraphics[width=0.45\linewidth]{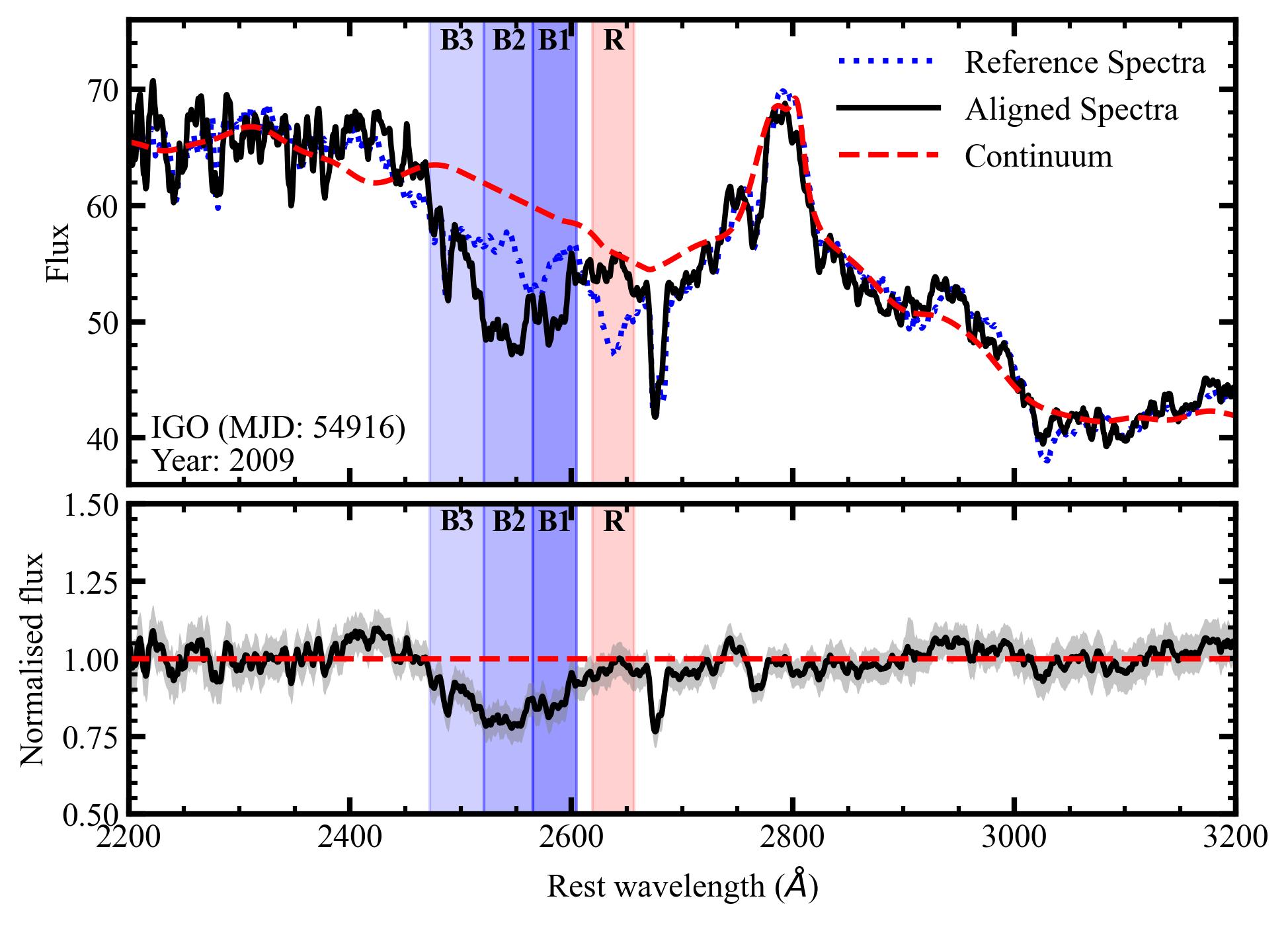}  & \includegraphics[width=0.45\linewidth]{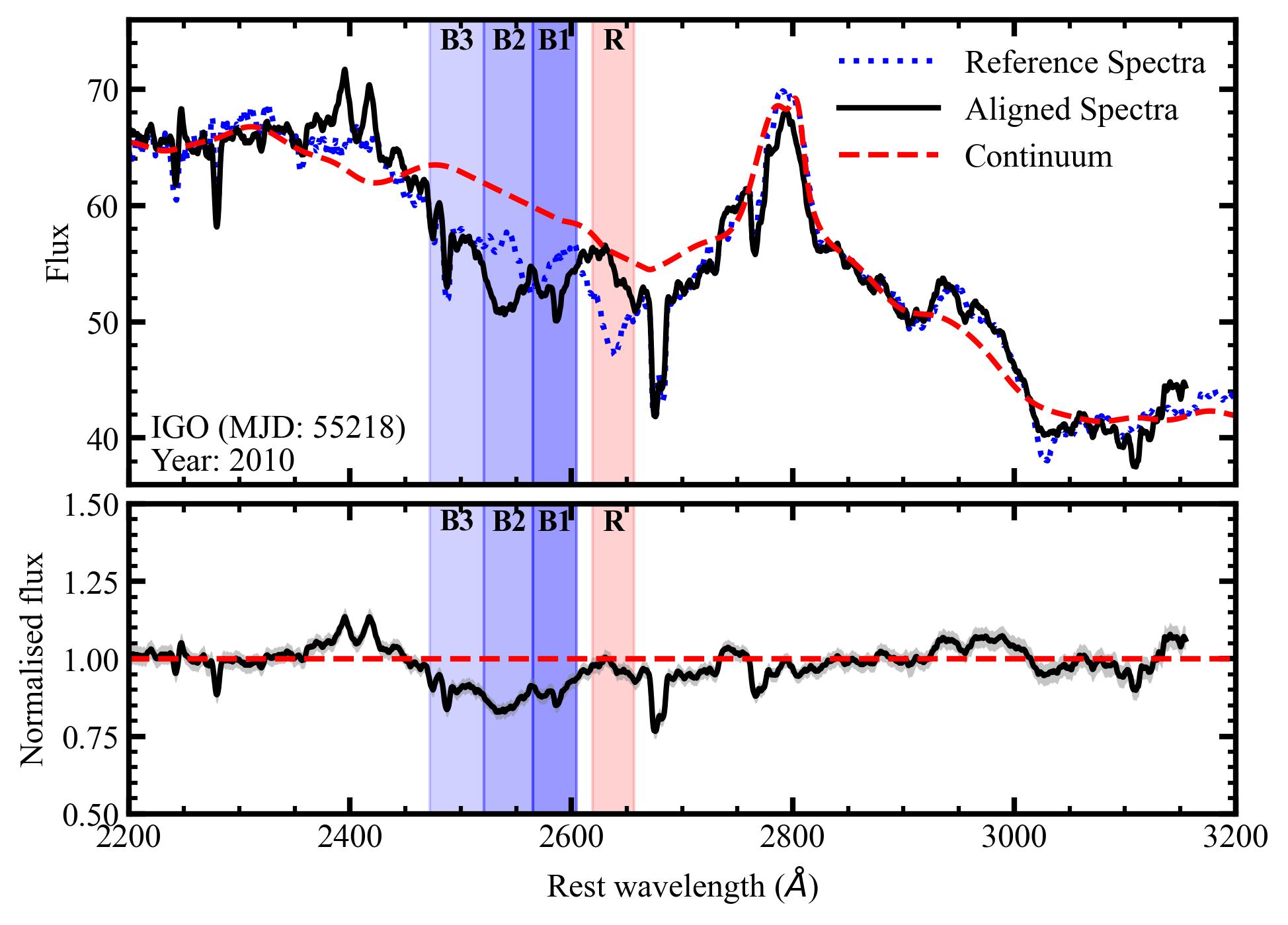} \\
             \includegraphics[width=0.45\linewidth]{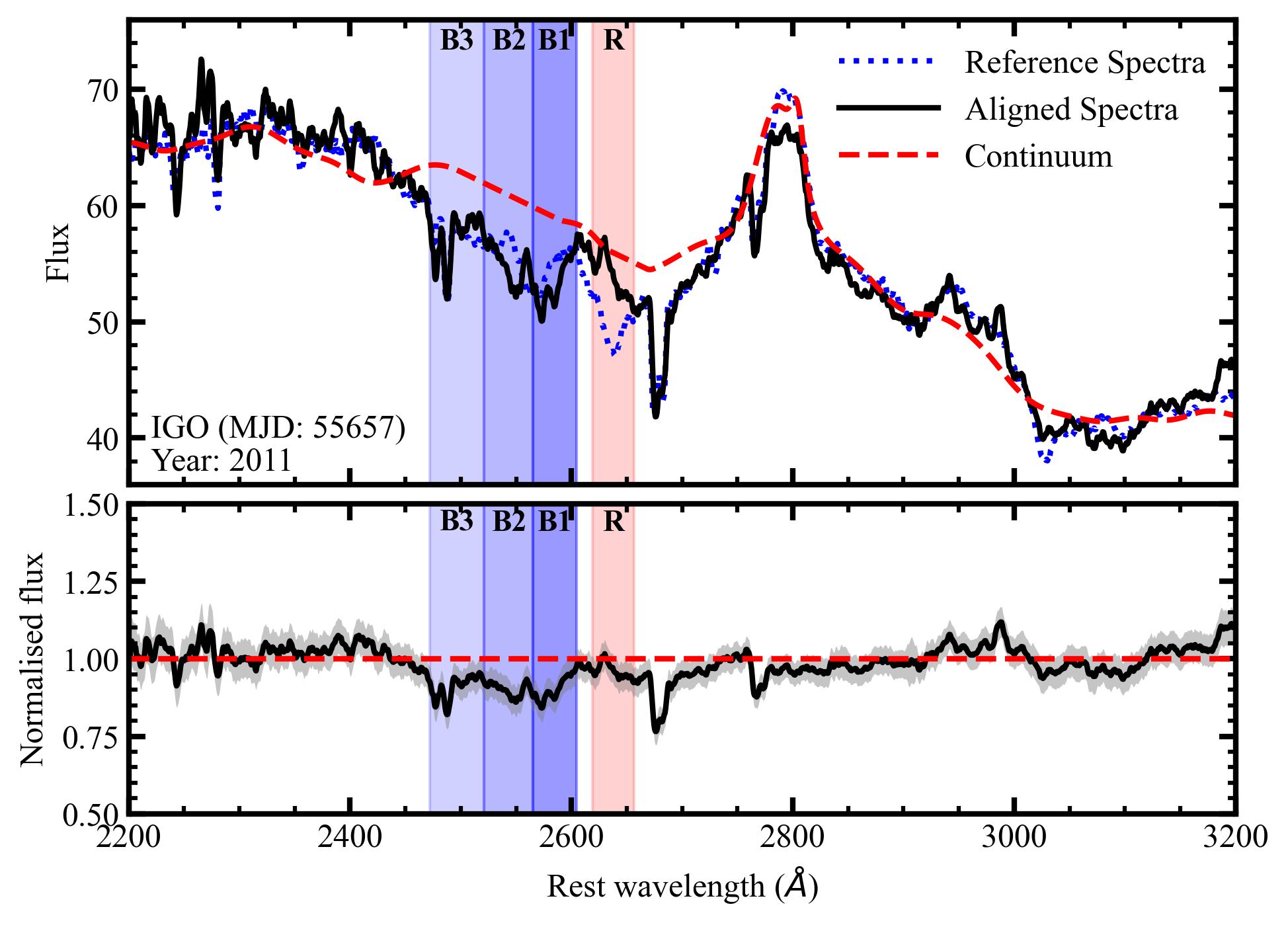}  & \includegraphics[width=0.45\linewidth]{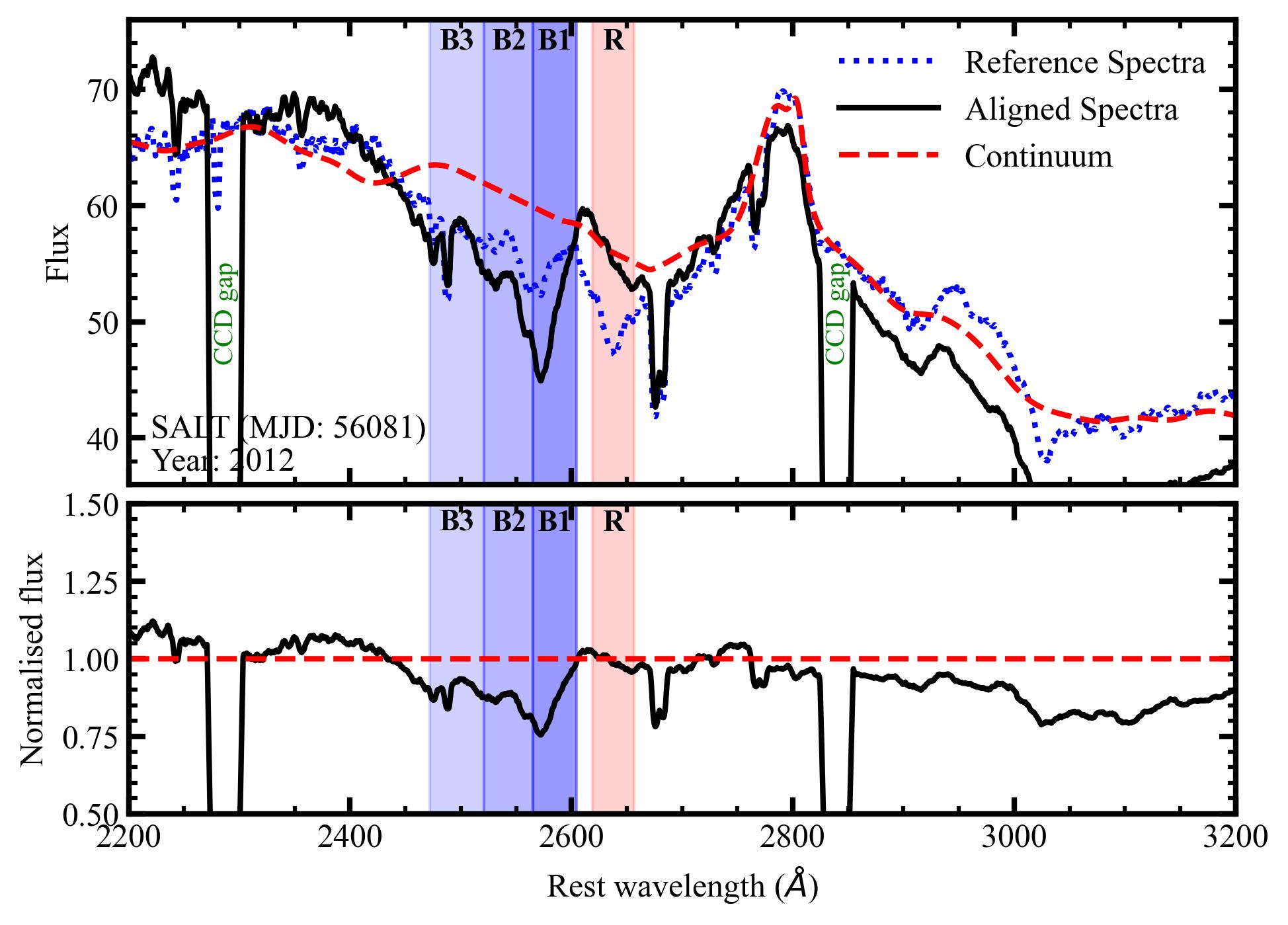} \\
        \end{tabular}
    \end{figure*}
     \begin{figure*}
        \begin{tabular}{ll}
           \includegraphics[width=0.45\linewidth]{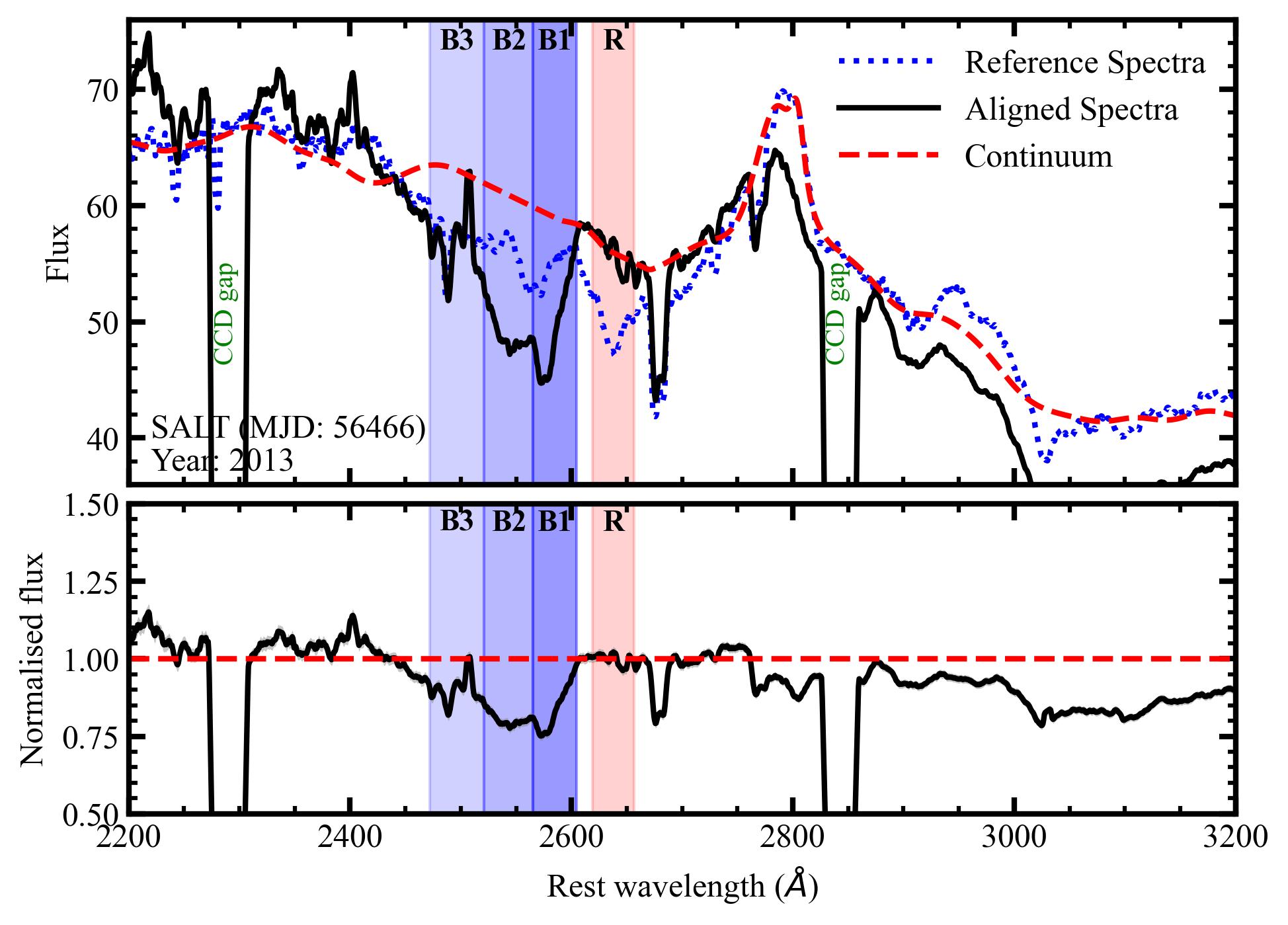}  & \includegraphics[width=0.45\linewidth]{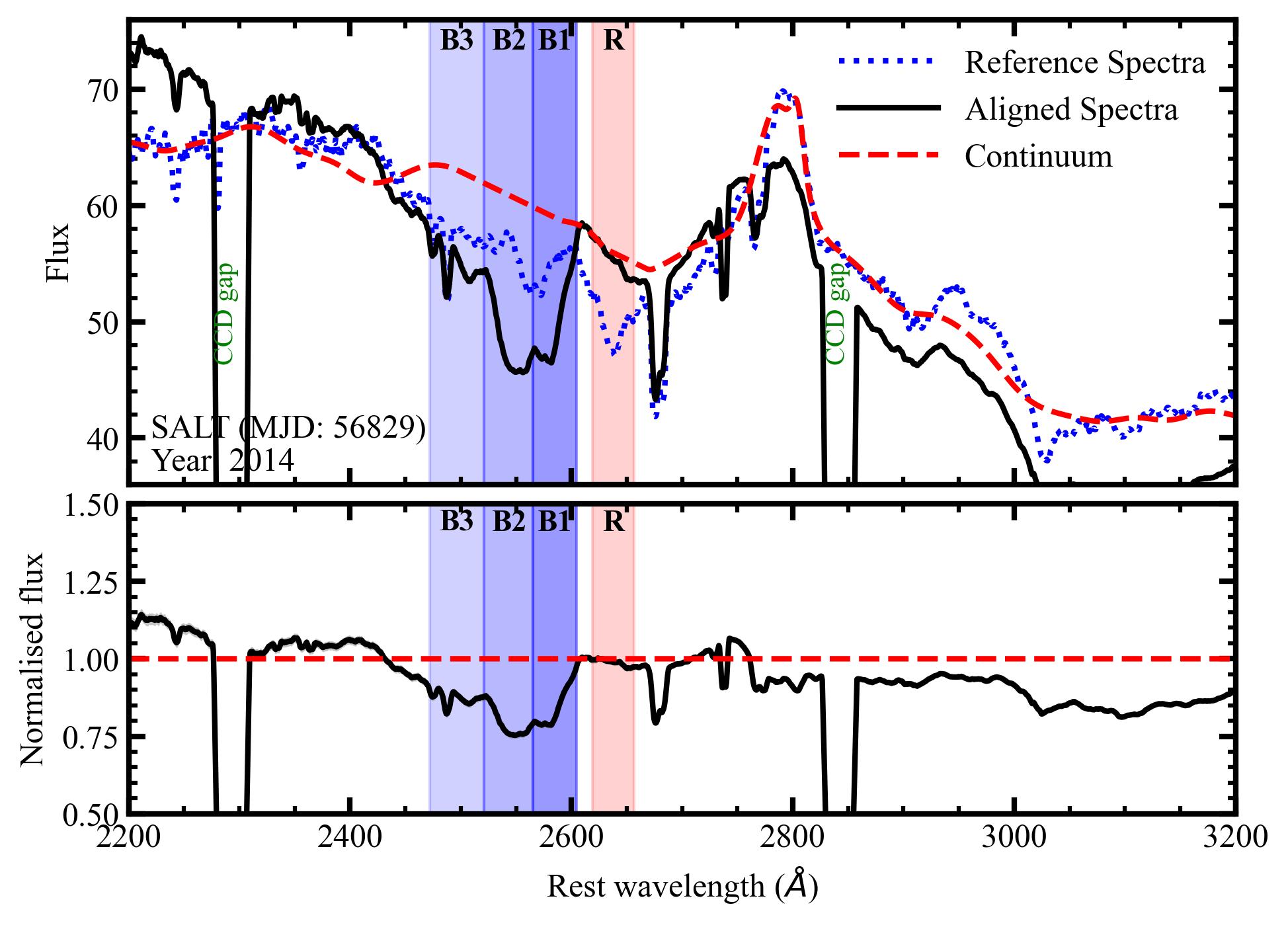} \\
             \includegraphics[width=0.45\linewidth]{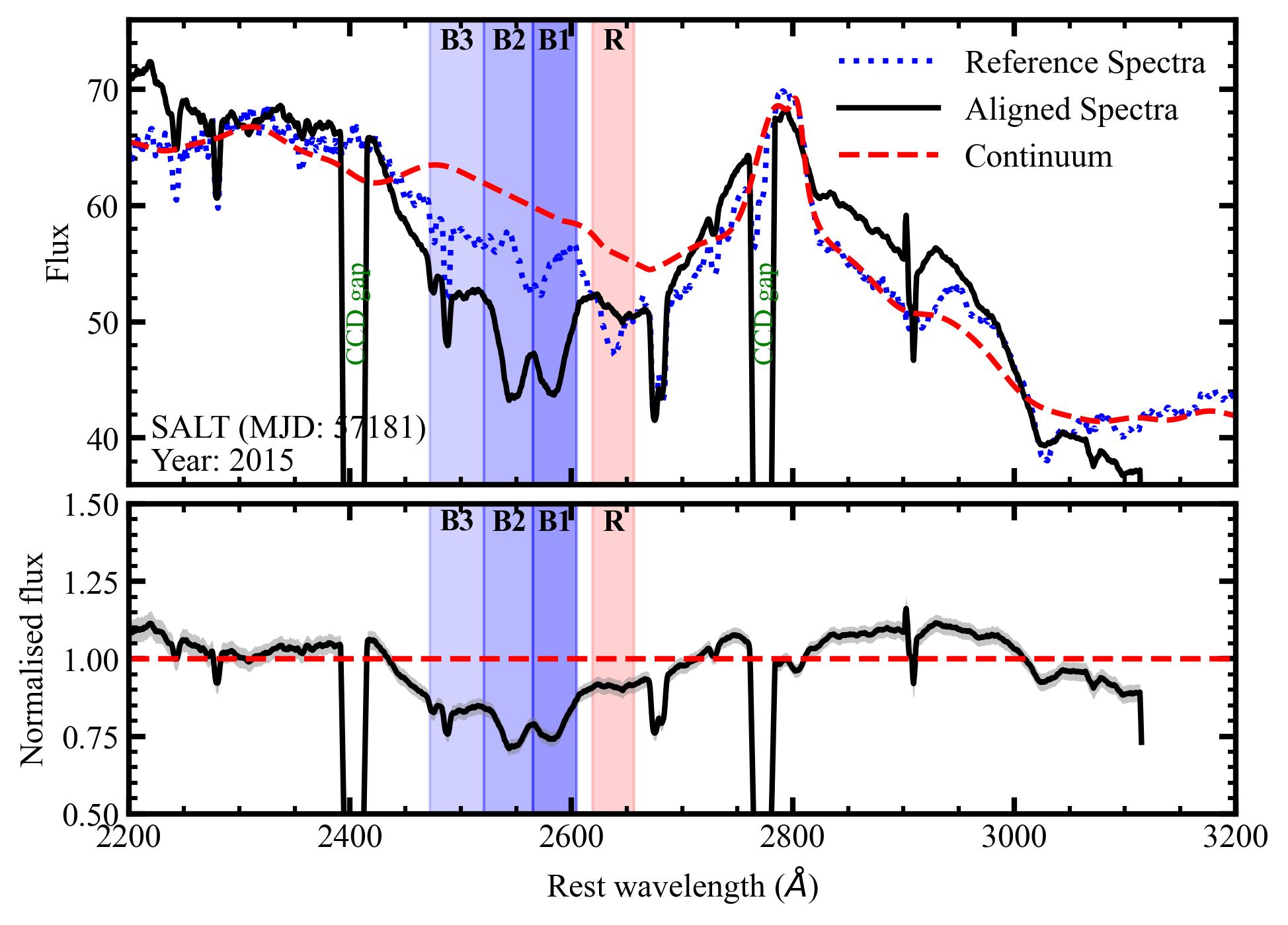}  & \includegraphics[width=0.45\linewidth]{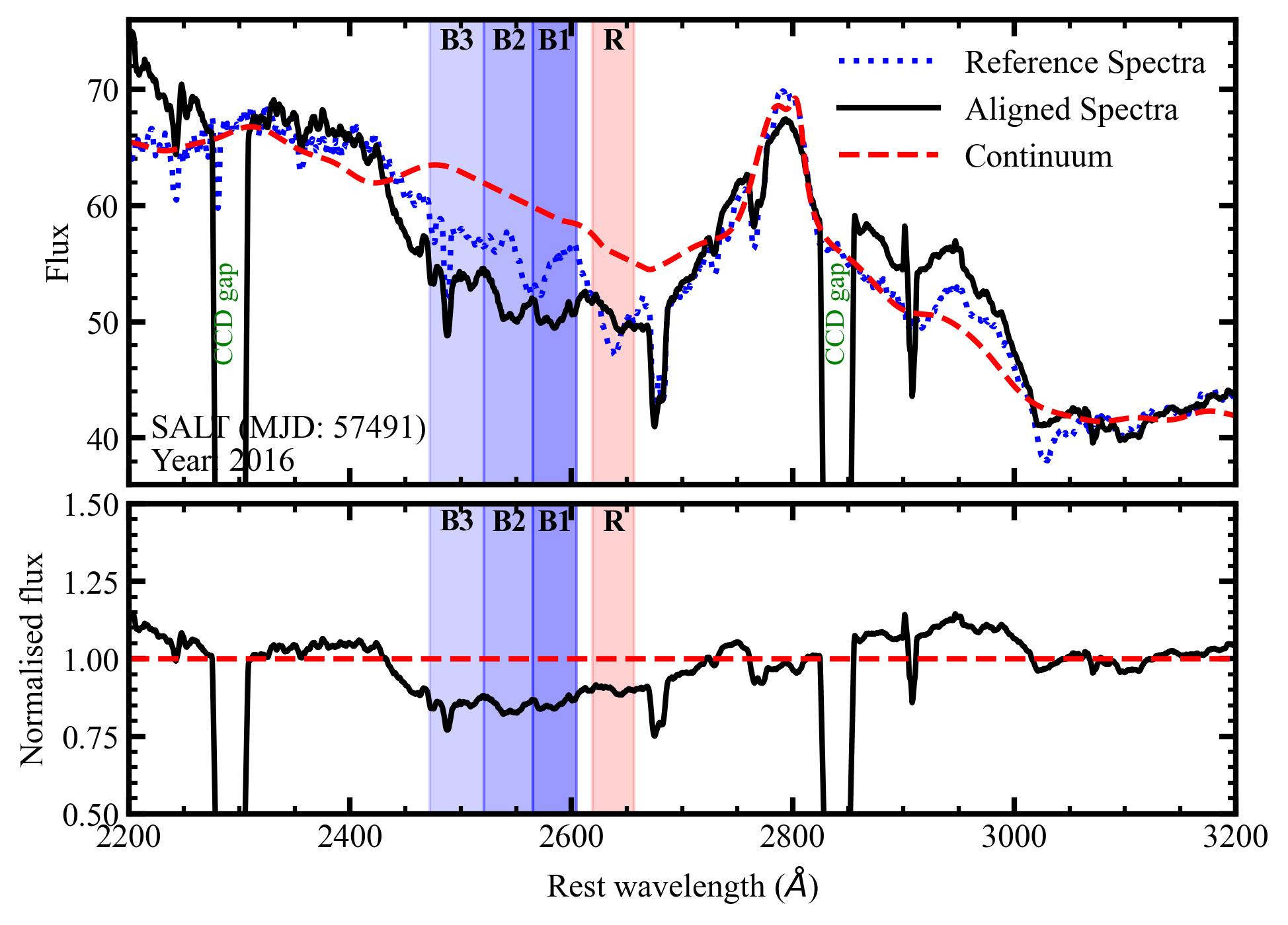} \\
             \includegraphics[width=0.45\linewidth]{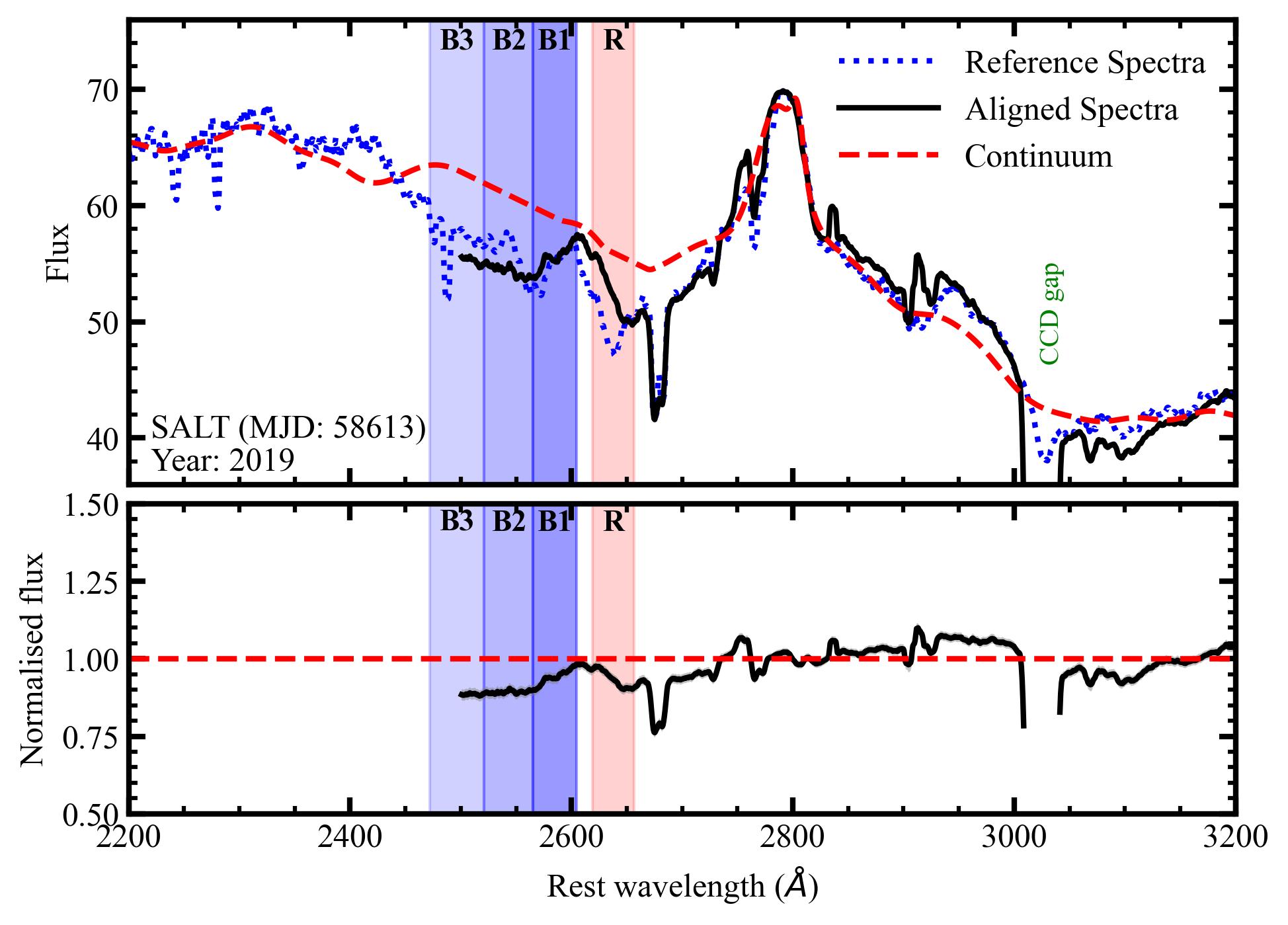}  & \includegraphics[width=0.45\linewidth]{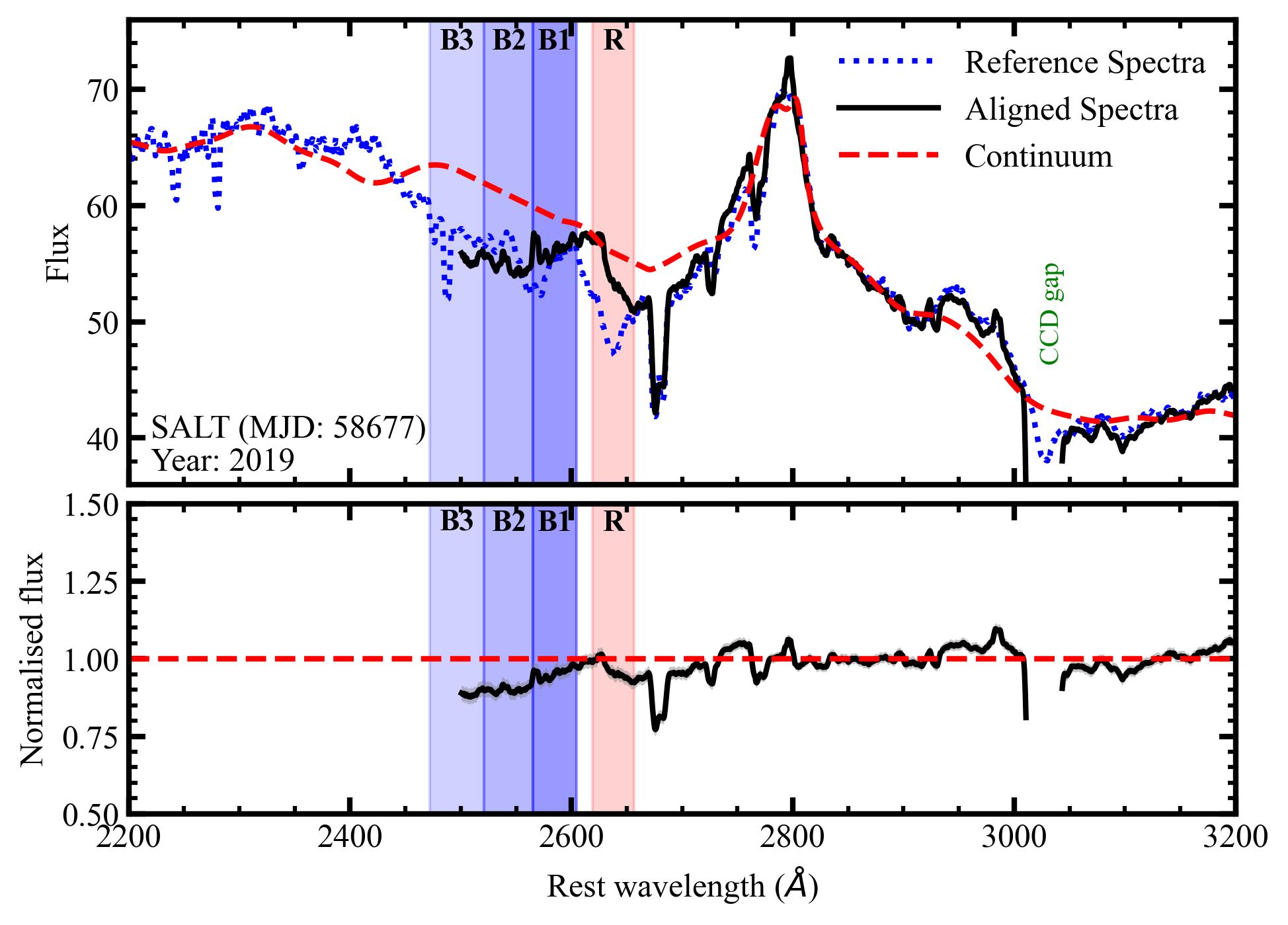}\\
             \includegraphics[width=0.45\linewidth]{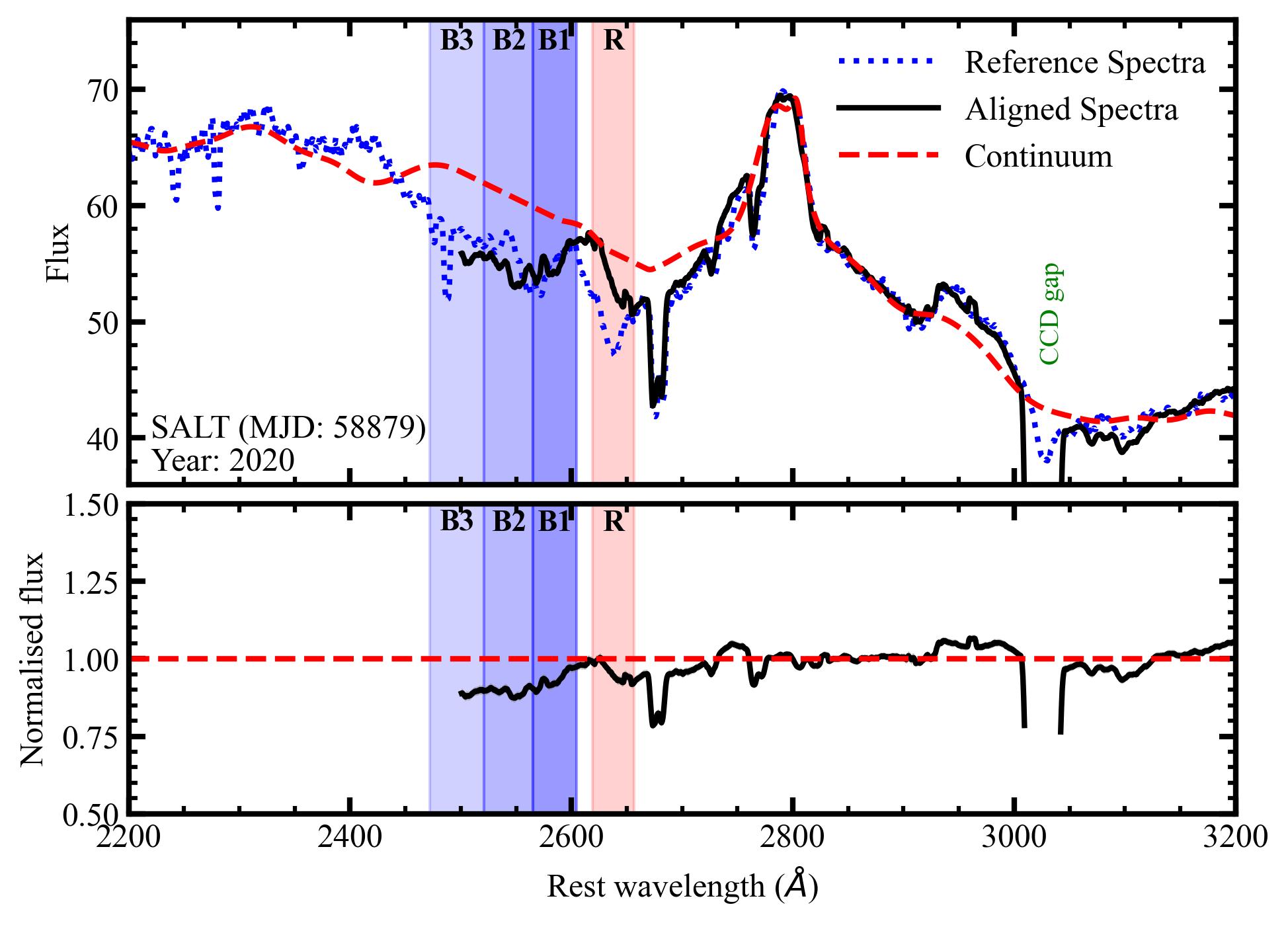}  & \includegraphics[width=0.45\linewidth]{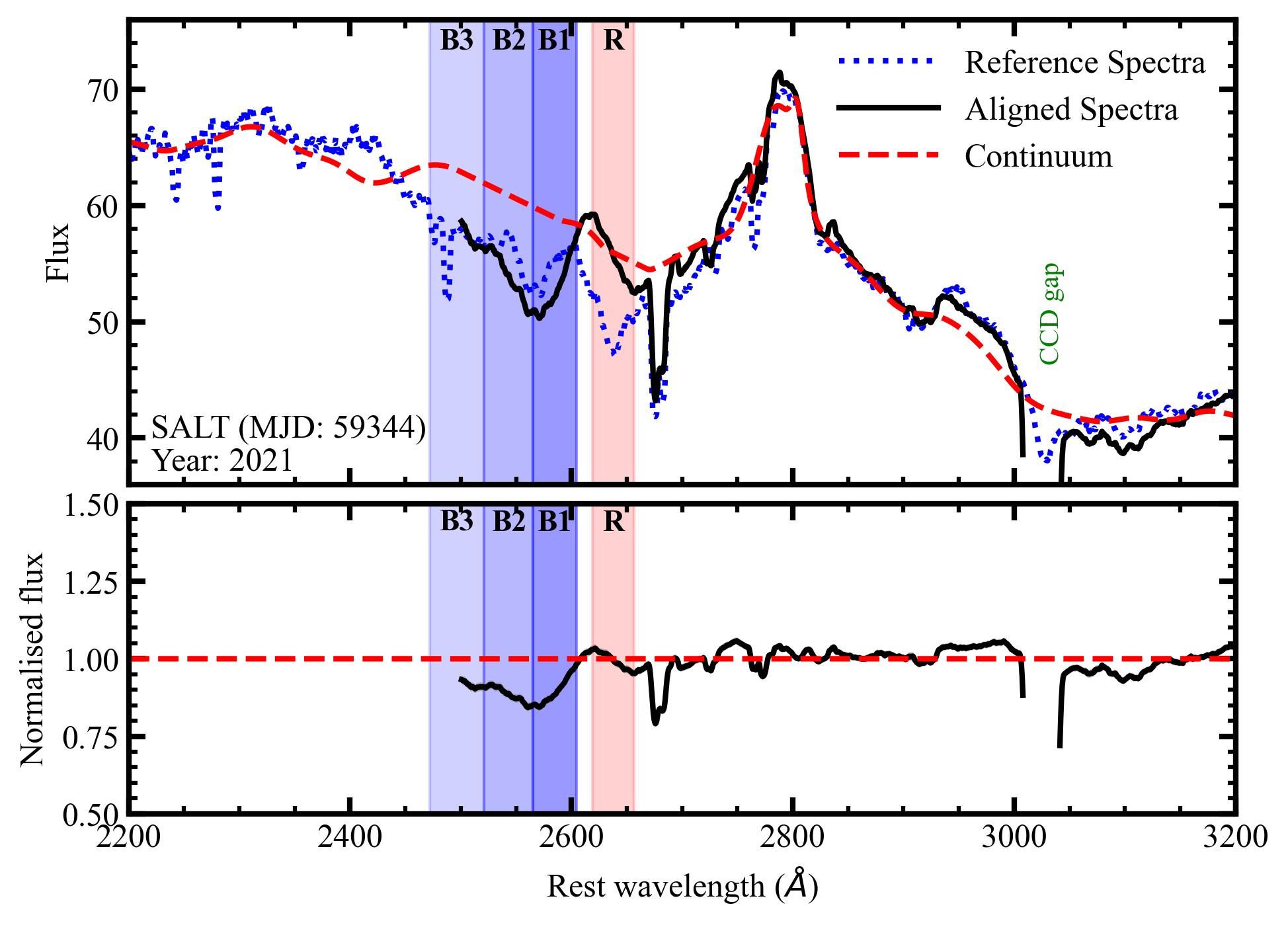} \\
        \end{tabular}
    \end{figure*}

    \begin{figure*}
        \begin{tabular}{ll}
        \includegraphics[width=0.45\linewidth]{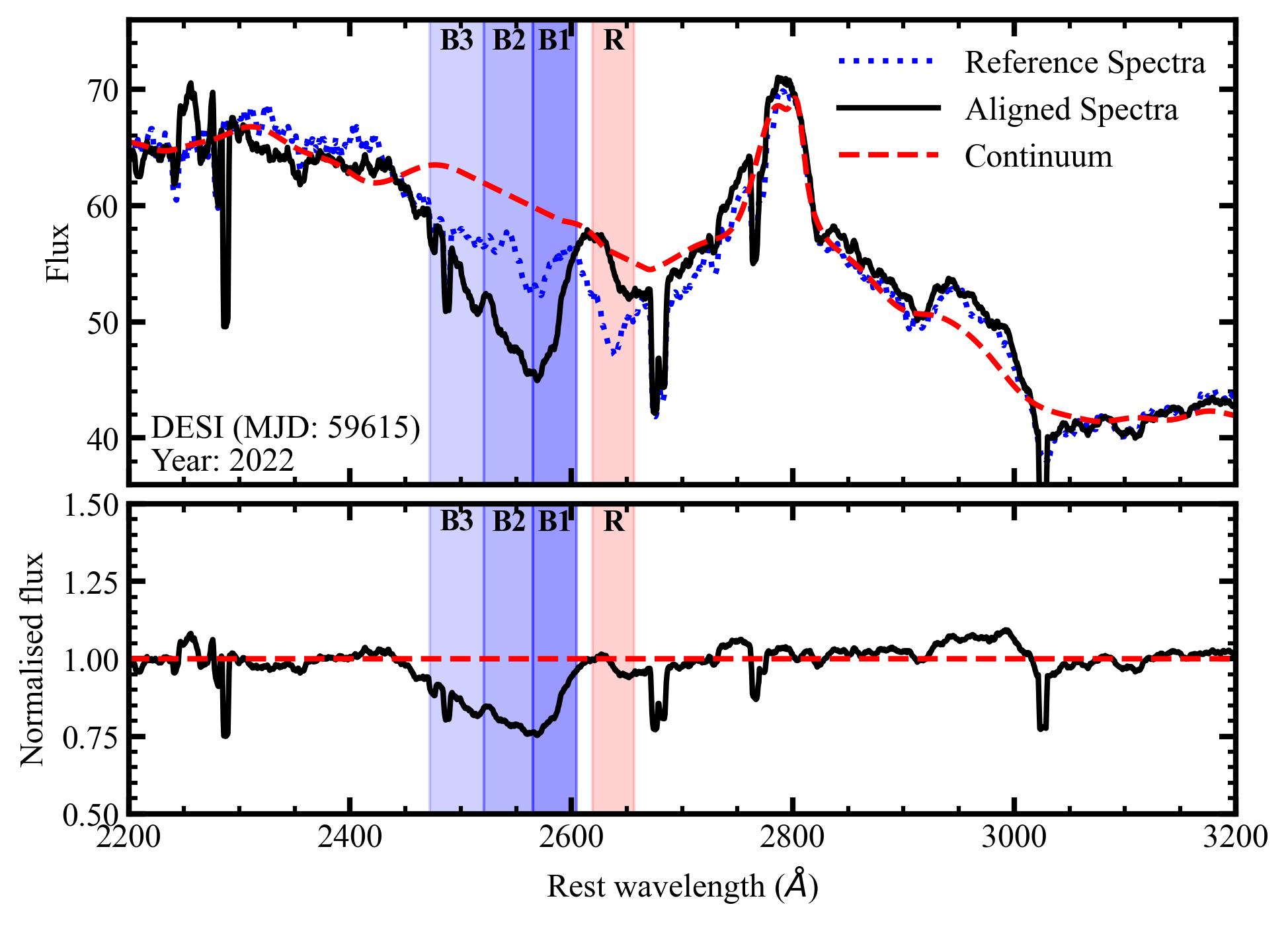} &
           \includegraphics[width=0.45\linewidth]{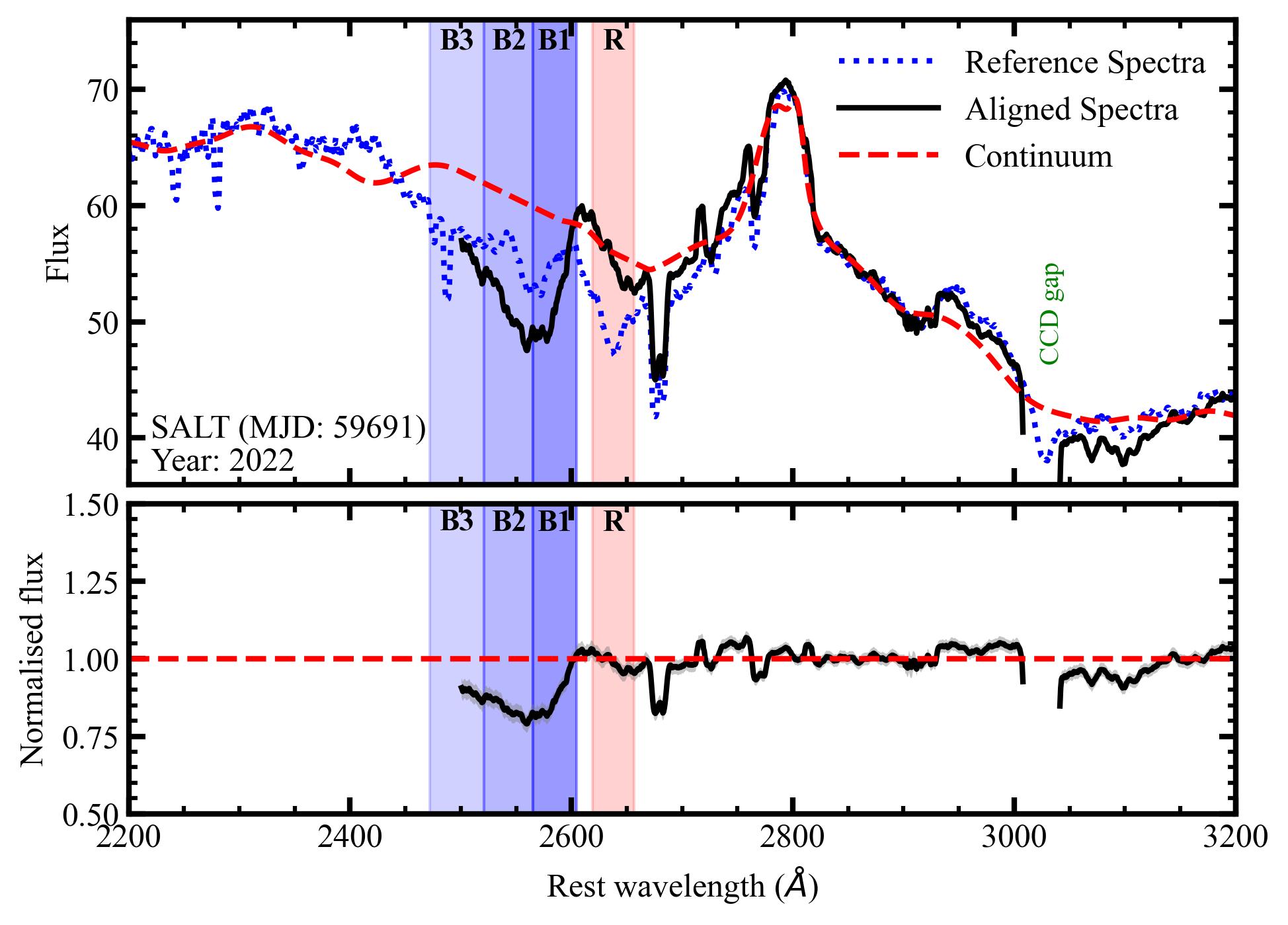}  \\ 
           \includegraphics[width=0.45\linewidth]{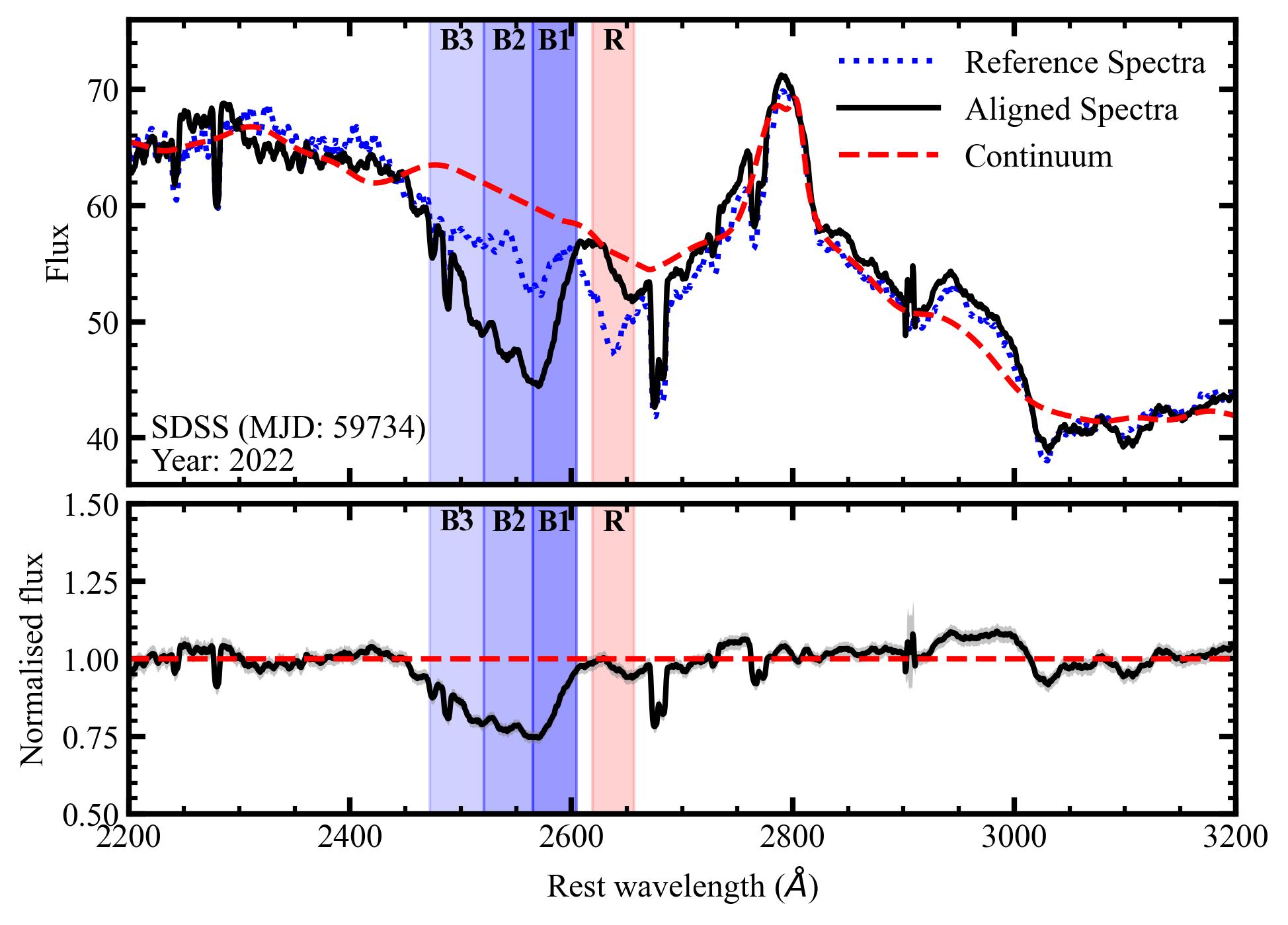} &\includegraphics[width=0.45\linewidth]{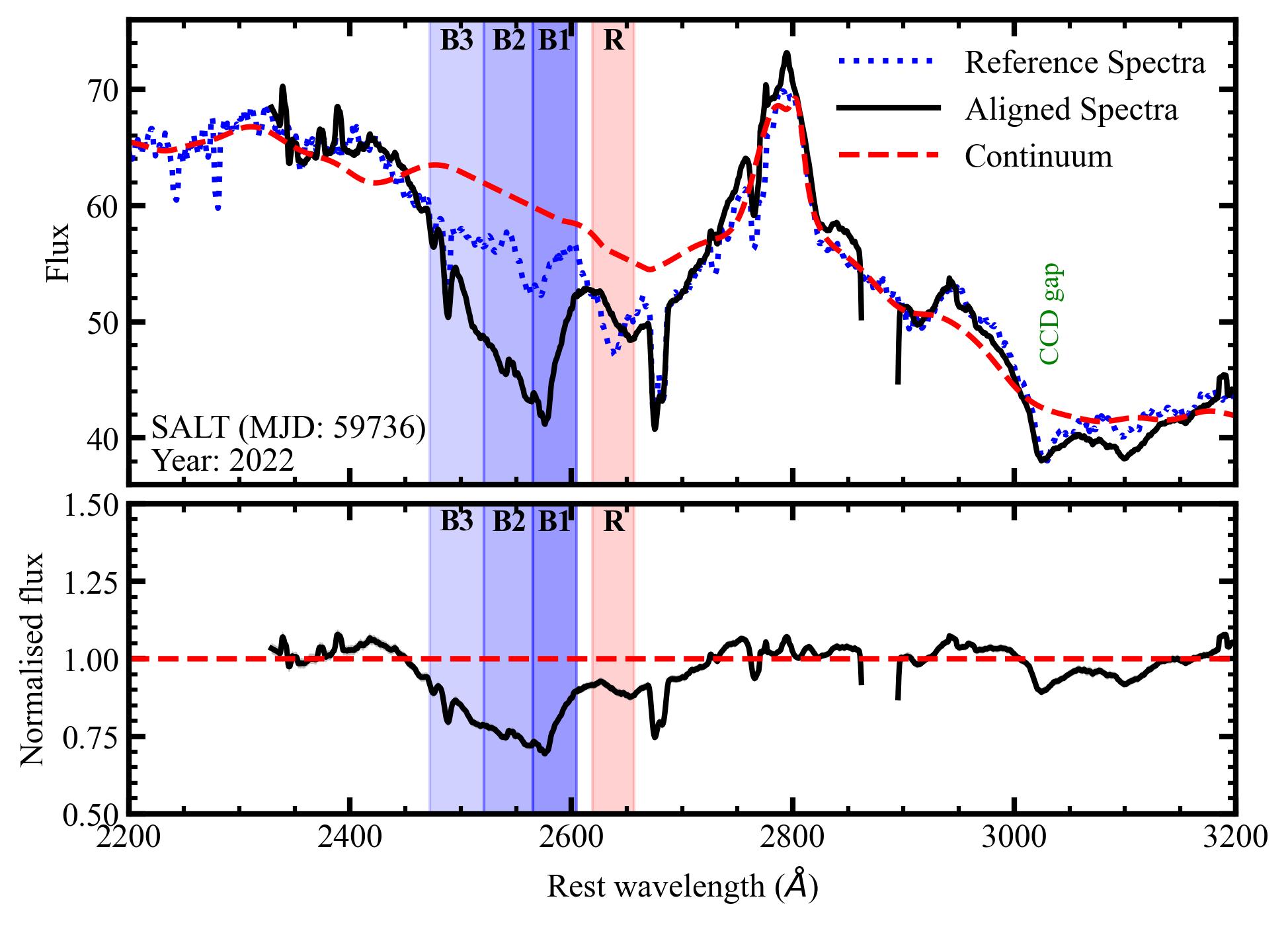} \\
           \includegraphics[width=0.45\linewidth]{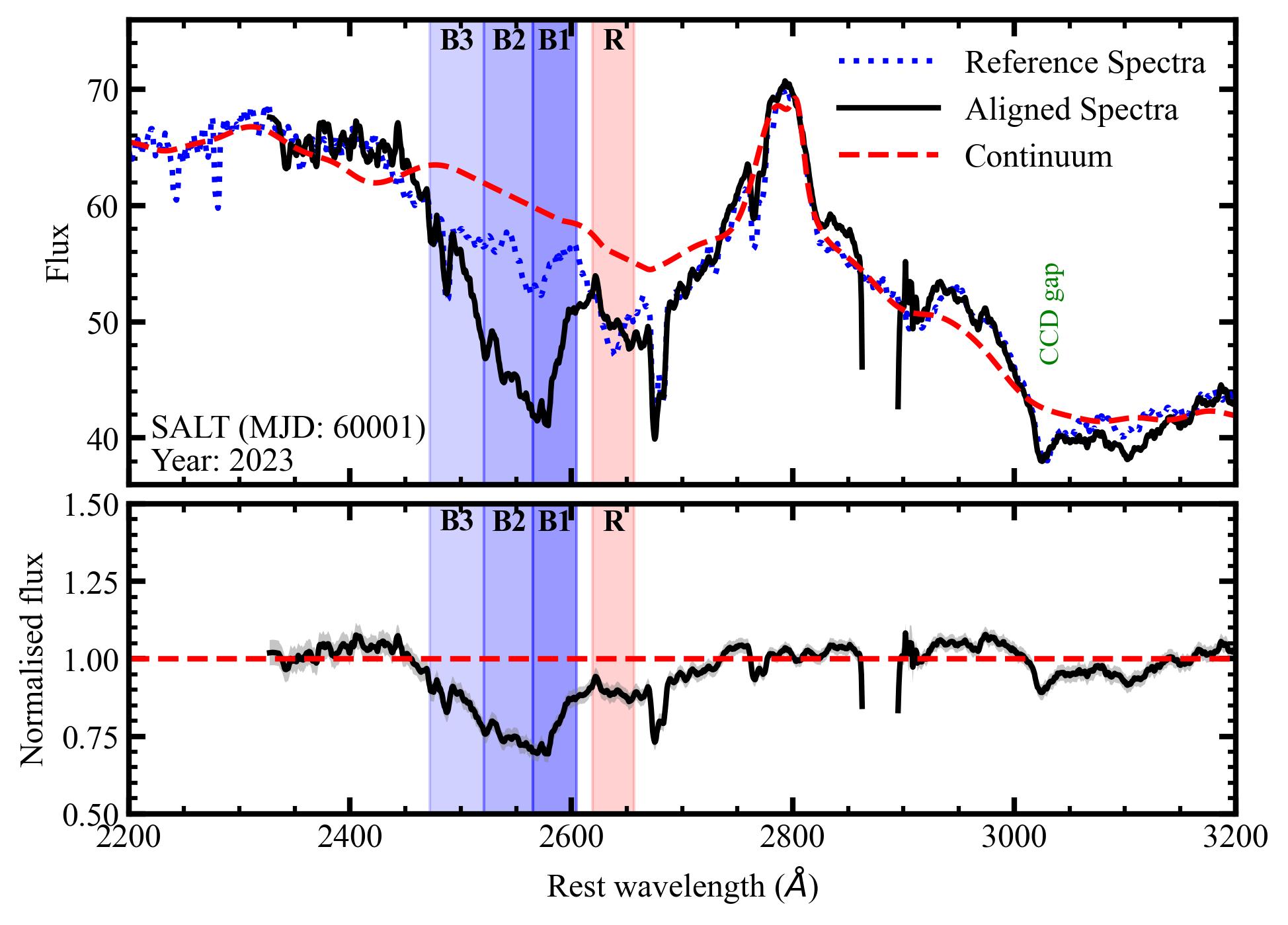} &\includegraphics[width=0.45\linewidth]{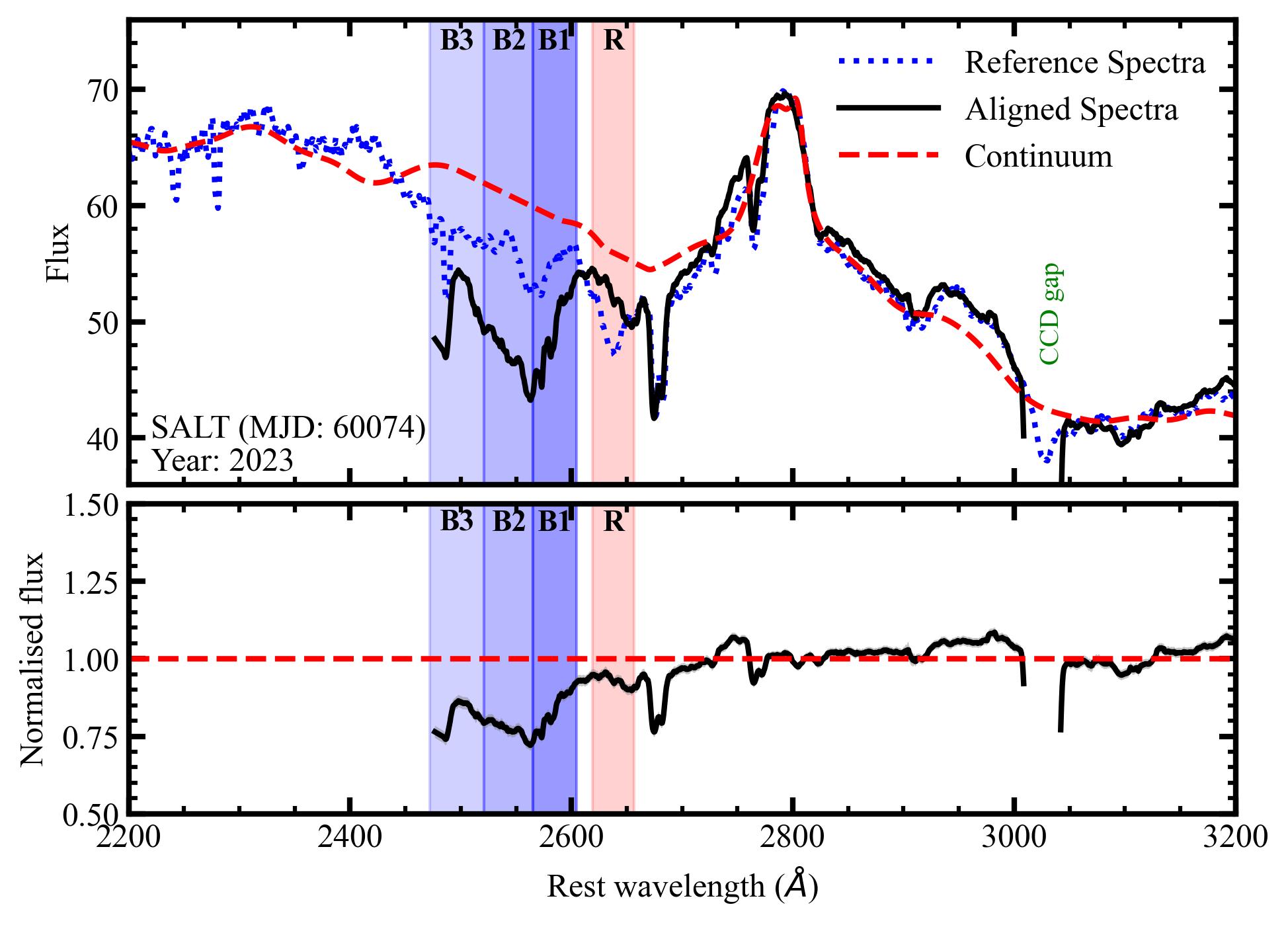} \\
           \includegraphics[width=0.45\linewidth]{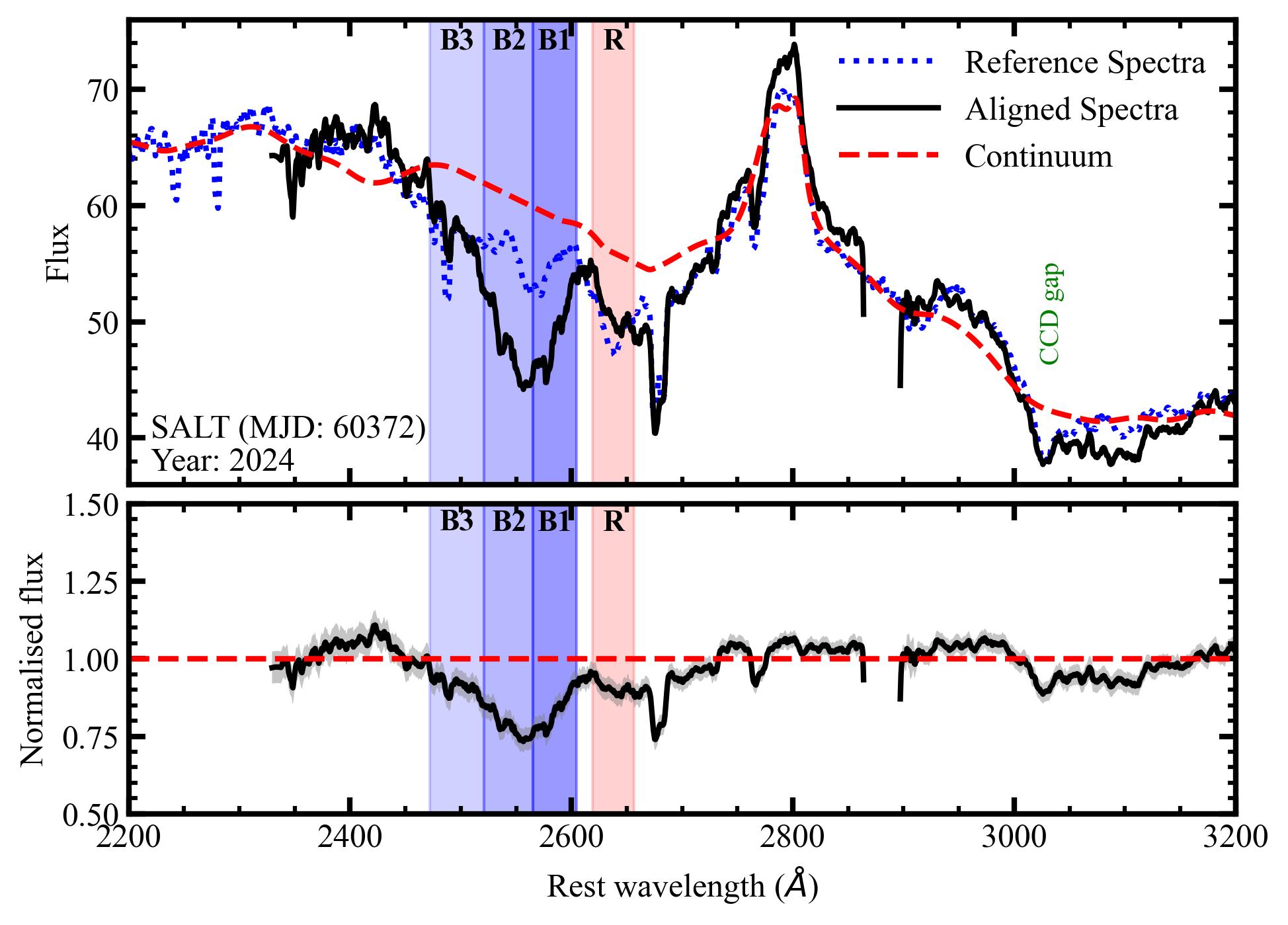} &\includegraphics[width=0.45\linewidth]{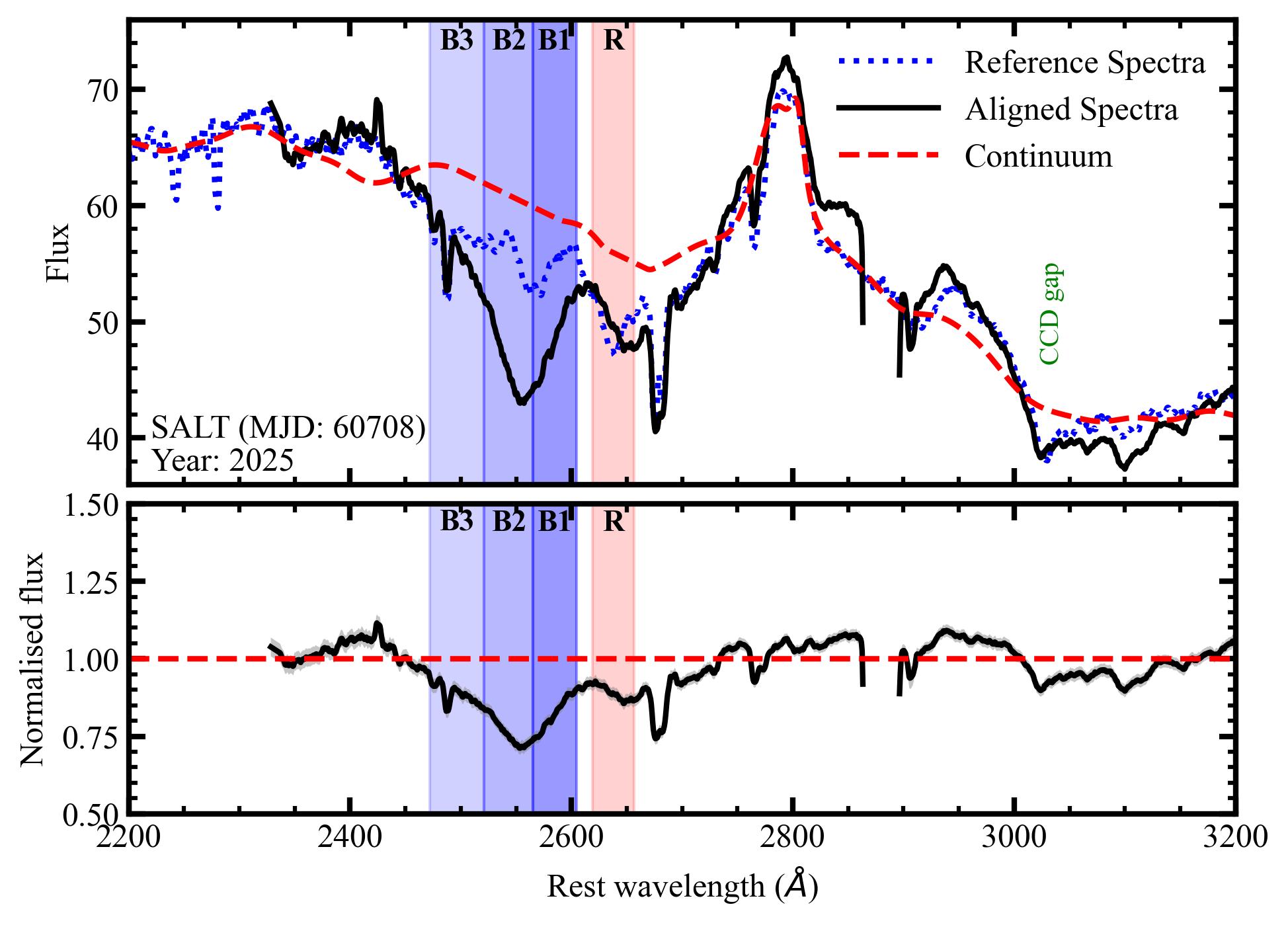} \\
             
        \end{tabular}
        \caption{Rest-frame spectra of J1333+0012 shown chronologically. In each panel, the observed spectrum is aligned to the reference SDSS spectrum at MJD 51955, with the fitted continuum overlaid. The lower panels show the continuum-normalized spectra with 1$\sigma$ uncertainties. Shaded regions mark the \mgii\ BAL components B3, B2, B1, and R. Narrow absorption redward of the R component is associated with an intervening absorber at $z_{\rm abs}\sim0.898$.}
       % \caption{Rest-frame spectra of J1333+0012 from different epochs, ordered chronologically. In each figure, the top panel shows the spectrum at a given epoch (black solid line) aligned to the reference SDSS spectrum obtained at MJD 51955 (blue dotted line), with the fitted continuum overlaid (red dashed line). The bottom panel presents the continuum-normalized spectrum, where the shaded grey region represents the one-sigma uncertainty. The shaded vertical bands mark the \mgii\ broad absorption line components: B3, B2, and B1 (blue) and R (red). Narrow absorption lines redward of the R-component are associated with the intervening absorbers at z$_{abs}$$\sim$0.898}
        \label{fig_continuum_plots}
    \end{figure*}

    \begin{figure*}
        \includegraphics[width=1\linewidth,trim=36 0 45 0, clip]{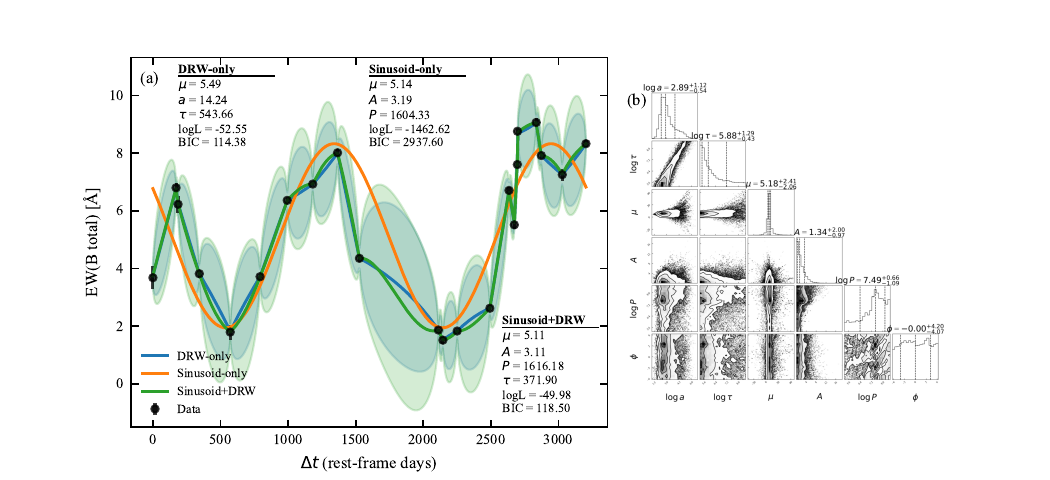}
        \caption{Same as Fig.~\ref{fig:periodicity}, but for B2 component alone. }
        \label{fig:periodicity_B2}
    \end{figure*}
\end{appendix}

%% For this sample we use BibTeX plus aasjournalv7.bst to generate the
%% the bibliography. The sample7.bib file was populated from ADS. To
%% get the citations to show in the compiled file do the following:
%%
%% pdflatex sample7.tex
%% bibtext sample7
%% pdflatex sample7.tex
%% pdflatex sample7.tex

\bibliography{j1333_bib}{}

@ARTICLE{Fabian2012,
       author = {{Fabian}, A.~C.},
        title = "{Observational Evidence of Active Galactic Nuclei Feedback}",
      journal = {\araa},
         year = 2012,
        month = sep,
       volume = {50},
        pages = {455-489},
          doi = {10.1146/annurev-astro-081811-125521},
archivePrefix = {arXiv},
       eprint = {1204.4114},
 primaryClass = {astro-ph.CO},
       adsurl = {https://ui.adsabs.harvard.edu/abs/2012ARA&A..50..455F}
}

@ARTICLE{Krolik1995,
       author = {{Krolik}, Julian H. and {Kriss}, Gerard A.},
        title = "{Observable Properties of X-Ray--heated Winds in Active Galactic Nuclei: Warm Reflectors and Warm Absorbers}",
      journal = {\apj},
         year = 1995,
        month = jul,
       volume = {447},
        pages = {512},
          doi = {10.1086/175896},
archivePrefix = {arXiv},
       eprint = {astro-ph/9501089},
 primaryClass = {astro-ph},
       adsurl = {https://ui.adsabs.harvard.edu/abs/1995ApJ...447..512K}
}

@ARTICLE{He2019,
       author = {{He}, Zhicheng and {Wang}, Tinggui and {Liu}, Guilin and {Wang}, Huiyuan and {Bian}, Weihao and {Tchernyshyov}, Kirill and {Mou}, Guobin and {Xu}, Youhua and {Zhou}, Hongyan and {Green}, Richard and {Xu}, Jun},
        title = "{The properties of broad absorption line outflows based on a large sample of quasars}",
      journal = {Nature Astronomy},
         year = 2019,
        month = jan,
       volume = {3},
        pages = {265},
          doi = {10.1038/s41550-018-0669-8},
archivePrefix = {arXiv},
       eprint = {1812.08982},
 primaryClass = {astro-ph.GA},
       adsurl = {https://ui.adsabs.harvard.edu/abs/2019NatAs...3..265H}
}

@ARTICLE{Kormendy2013,
       author = {{Kormendy}, John and {Ho}, Luis C.},
        title = "{Coevolution (Or Not) of Supermassive Black Holes and Host Galaxies}",
      journal = {\araa},
         year = 2013,
        month = aug,
       volume = {51},
       number = {1},
        pages = {511-653},
          doi = {10.1146/annurev-astro-082708-101811},
archivePrefix = {arXiv},
       eprint = {1304.7762},
 primaryClass = {astro-ph.CO},
       adsurl = {https://ui.adsabs.harvard.edu/abs/2013ARA&A..51..511K}
}

@ARTICLE{Hiremath2025,
       author = {{Hiremath}, Pranavi and {Rankine}, Amy L. and {Aird}, James and {Brandt}, W.~N. and {Rodr{\'\i}guez Hidalgo}, Paola and {Anderson}, Scott F. and {Aydar}, Catarina and {Ricci}, Claudio and {Schneider}, Donald P. and {Vivek}, M. and {Igo}, Zsofi and {Morrison}, Sean and {Salvato}, Mara},
        title = "{X-ray selected broad absorption line quasars in SDSS-V: BALs and non-BALs span the same range of X-ray properties}",
      journal = {\mnras},
         year = 2025,
        month = sep,
       volume = {542},
       number = {3},
        pages = {2105-2127},
          doi = {10.1093/mnras/staf1352},
archivePrefix = {arXiv},
       eprint = {2508.13682},
 primaryClass = {astro-ph.GA},
       adsurl = {https://ui.adsabs.harvard.edu/abs/2025MNRAS.542.2105H}
}

@ARTICLE{Hall2002,
       author = {{Hall}, Patrick B. and {Anderson}, Scott F. and {Strauss}, Michael A. and {York}, Donald G. and {Richards}, Gordon T. and {Fan}, Xiaohui and {Knapp}, G.~R. and {Schneider}, Donald P. and {Vanden Berk}, Daniel E. and {Geballe}, T.~R. and {Bauer}, Amanda E. and {Becker}, Robert H. and {Davis}, Marc and {Rix}, Hans-Walter and {Nichol}, R.~C. and {Bahcall}, Neta A. and {Brinkmann}, J. and {Brunner}, Robert and {Connolly}, A.~J. and {Csabai}, Istv{\'a}n and {Doi}, Mamoru and {Fukugita}, Masataka and {Gunn}, James E. and {Haiman}, Zoltan and {Harvanek}, Michael and {Heckman}, Timothy M. and {Hennessy}, G.~S. and {Inada}, Naohisa and {Ivezi{\'c}}, {\v{Z}}eljko and {Johnston}, David and {Kleinman}, S. and {Krolik}, Julian H. and {Krzesinski}, Jurek and {Kunszt}, Peter Z. and {Lamb}, D.~Q. and {Long}, Daniel C. and {Lupton}, Robert H. and {Miknaitis}, Gajus and {Munn}, Jeffrey A. and {Narayanan}, Vijay K. and {Neilsen}, Eric and {Newman}, P.~R. and {Nitta}, Atsuko and {Okamura}, Sadanori and {Pentericci}, Laura and {Pier}, Jeffrey R. and {Schlegel}, David J. and {Snedden}, S. and {Szalay}, Alexander S. and {Thakar}, Anirudda R. and {Tsvetanov}, Zlatan and {White}, Richard L. and {Zheng}, Wei},
        title = "{Unusual Broad Absorption Line Quasars from the Sloan Digital Sky Survey}",
      journal = {\apjs},
         year = 2002,
        month = aug,
       volume = {141},
       number = {2},
        pages = {267-309},
          doi = {10.1086/340546},
archivePrefix = {arXiv},
       eprint = {astro-ph/0203252},
 primaryClass = {astro-ph},
       adsurl = {https://ui.adsabs.harvard.edu/abs/2002ApJS..141..267H}
}

@ARTICLE{vivek14,
   author = {{Vivek}, M. and {Srianand}, R. and {Petitjean}, P. and {Mohan}, V. and 
	{Mahabal}, A. and {Samui}, S.},
    title = "{Variability in Low Ionization Broad Absorption Line outflows}",
  journal = {\mnras},
archivePrefix = "arXiv",
   eprint = {1402.2980},
 primaryClass = "astro-ph.CO",
     year = 2014,
    month = may,
   volume = 440,
    pages = {799-820},
      doi = {10.1093/mnras/stu288},
   adsurl = {http://adsabs.harvard.edu/abs/2014MNRAS.440..799V}
}

@ARTICLE{vivek12,
   author = {{Vivek}, M. and {Srianand}, R. and {Mahabal}, A. and {Kuriakose}, V.~C.
	},
    title = "{Dynamically evolving Mg II broad absorption line flow in SDSS J133356.02+001229.1}",
  journal = {\mnras},
archivePrefix = "arXiv",
   eprint = {1201.0431},
 primaryClass = "astro-ph.CO",
     year = 2012,
    month = mar,
   volume = 421,
    pages = {L107-L111},
      doi = {10.1111/j.1745-3933.2012.01216.x},
   adsurl = {http://adsabs.harvard.edu/abs/2012MNRAS.421L.107V}
}

@ARTICLE{weynman91,
   author = {{Weymann}, R.~J. and {Morris}, S.~L. and {Foltz}, C.~B. and 
	{Hewett}, P.~C.},
    title = "{Comparisons of the emission-line and continuum properties of broad absorption line and normal quasi-stellar objects}",
  journal = {\apj},
     year = 1991,
    month = may,
   volume = 373,
    pages = {23-53},
      doi = {10.1086/170020},
   adsurl = {http://adsabs.harvard.edu/abs/1991ApJ...373...23W}
}

@ARTICLE{filiz13,
   author = {{Filiz Ak}, N. and {Brandt}, W.~N. and {Hall}, P.~B. and {Schneider}, D.~P. and 
	{Anderson}, S.~F. and {Hamann}, F. and {Lundgren}, B.~F. and 
	{Myers}, A.~D. and {P{\^a}ris}, I. and {Petitjean}, P. and {Ross}, N.~P. and 
	{Shen}, Y. and {York}, D.},
    title = "{Broad Absorption Line Variability on Multi-year Timescales in a Large Quasar Sample}",
  journal = {\apj},
archivePrefix = "arXiv",
   eprint = {1309.5364},
     year = 2013,
    month = nov,
   volume = 777,
      eid = {168},
    pages = {168},
      doi = {10.1088/0004-637X/777/2/168},
   adsurl = {http://adsabs.harvard.edu/abs/2013ApJ...777..168F}
}

@ARTICLE{Aromal2025,
       author = {{Aromal}, P. and {Srianand}, R. and {Gallagher}, S.~C. and {Vivek}, M. and {Petitjean}, P.},
        title = "{Transient LoBALs at High Velocities: A Case of Extreme Broad Absorption Line Variability in J115636.82+085628.9}",
      journal = {\apj},
         year = 2025,
        month = sep,
       volume = {990},
       number = {2},
          eid = {146},
        pages = {146},
          doi = {10.3847/1538-4357/adf4de},
archivePrefix = {arXiv},
       eprint = {2508.01029},
 primaryClass = {astro-ph.GA},
       adsurl = {https://ui.adsabs.harvard.edu/abs/2025ApJ...990..146A}
}

@ARTICLE{vivek2018,
       author = {{Vivek}, M. and {Srianand}, R. and {Dawson}, K.~S.},
        title = "{Rapidly varying Mg II broad absorption line in SDSS J133356.02 + 001229.1}",
      journal = {\mnras},
         year = 2018,
        month = dec,
       volume = {481},
       number = {4},
        pages = {5570-5579},
          doi = {10.1093/mnras/sty2712},
archivePrefix = {arXiv},
       eprint = {1809.10155},
 primaryClass = {astro-ph.GA},
       adsurl = {https://ui.adsabs.harvard.edu/abs/2018MNRAS.481.5570V}
}

@ARTICLE{Shen2024,
       author = {{Shen}, Yue and {Grier}, Catherine J. and {Horne}, Keith and {Stone}, Zachary and {Li}, Jennifer I. and {Yang}, Qian and {Homayouni}, Yasaman and {Trump}, Jonathan R. and {Anderson}, Scott F. and {Brandt}, W.~N. and {Hall}, Patrick B. and {Ho}, Luis C. and {Jiang}, Linhua and {Petitjean}, Patrick and {Schneider}, Donald P. and {Tao}, Charling and {Donnan}, Fergus. R. and {AlSayyad}, Yusra and {Bershady}, Matthew A. and {Blanton}, Michael R. and {Bizyaev}, Dmitry and {Bundy}, Kevin and {Chen}, Yuguang and {Davis}, Megan C. and {Dawson}, Kyle and {Fan}, Xiaohui and {Greene}, Jenny E. and {Gr{\"o}ller}, Hannes and {Guo}, Yucheng and {Ibarra-Medel}, H{\'e}ctor and {Jiang}, Yuanzhe and {Keenan}, Ryan P. and {Kollmeier}, Juna A. and {Lejoly}, Cassandra and {Li}, Zefeng and {de la Macorra}, Axel and {Moe}, Maxwell and {Nie}, Jundan and {Rossi}, Graziano and {Smith}, Paul S. and {Tee}, Wei Leong and {Weijmans}, Anne-Marie and {Xu}, Jiachuan and {Yue}, Minghao and {Zhou}, Xu and {Zhou}, Zhimin and {Zou}, Hu},
        title = "{The Sloan Digital Sky Survey Reverberation Mapping Project: Key Results}",
      journal = {\apjs},
         year = 2024,
        month = jun,
       volume = {272},
       number = {2},
          eid = {26},
        pages = {26},
          doi = {10.3847/1538-4365/ad3936},
archivePrefix = {arXiv},
       eprint = {2305.01014},
 primaryClass = {astro-ph.GA},
       adsurl = {https://ui.adsabs.harvard.edu/abs/2024ApJS..272...26S}
}

@ARTICLE{Homayani2020,
       author = {{Homayouni}, Y. and {Trump}, Jonathan R. and {Grier}, C.~J. and {Horne}, Keith and {Shen}, Yue and {Brandt}, W.~N. and {Dawson}, Kyle S. and {Alvarez}, Gloria Fonseca and {Green}, Paul J. and {Hall}, P.~B. and {Hern{\'a}ndez Santisteban}, Juan V. and {Ho}, Luis C. and {Kinemuchi}, Karen and {Kochanek}, C.~S. and {Li}, Jennifer I.-Hsiu and {Peterson}, B.~M. and {Schneider}, D.~P. and {Starkey}, D.~A. and {Bizyaev}, Dmitry and {Pan}, Kaike and {Oravetz}, Daniel and {Simmons}, Audrey},
        title = "{The Sloan Digital Sky Survey Reverberation Mapping Project: Mg II Lag Results from Four Years of Monitoring}",
      journal = {\apj},
         year = 2020,
        month = sep,
       volume = {901},
       number = {1},
          eid = {55},
        pages = {55},
          doi = {10.3847/1538-4357/ababa9},
archivePrefix = {arXiv},
       eprint = {2005.03663},
 primaryClass = {astro-ph.GA},
       adsurl = {https://ui.adsabs.harvard.edu/abs/2020ApJ...901...55H}
}

@ARTICLE{Wu2022,
       author = {{Wu}, Qiaoya and {Shen}, Yue},
        title = "{A Catalog of Quasar Properties from Sloan Digital Sky Survey Data Release 16}",
      journal = {\apjs},
         year = 2022,
        month = dec,
       volume = {263},
       number = {2},
          eid = {42},
        pages = {42},
          doi = {10.3847/1538-4365/ac9ead},
archivePrefix = {arXiv},
       eprint = {2209.03987},
 primaryClass = {astro-ph.GA},
       adsurl = {https://ui.adsabs.harvard.edu/abs/2022ApJS..263...42W}
}

@ARTICLE{macleod2010,
       author = {{MacLeod}, C.~L. and {Ivezi{\'c}}, {\v{Z}}. and {Kochanek}, C.~S. and {Koz{\l}owski}, S. and {Kelly}, B. and {Bullock}, E. and {Kimball}, A. and {Sesar}, B. and {Westman}, D. and {Brooks}, K. and {Gibson}, R. and {Becker}, A.~C. and {de Vries}, W.~H.},
        title = "{Modeling the Time Variability of SDSS Stripe 82 Quasars as a Damped Random Walk}",
      journal = {\apj},
         year = 2010,
        month = oct,
       volume = {721},
       number = {2},
        pages = {1014-1033},
          doi = {10.1088/0004-637X/721/2/1014},
archivePrefix = {arXiv},
       eprint = {1004.0276},
 primaryClass = {astro-ph.CO},
       adsurl = {https://ui.adsabs.harvard.edu/abs/2010ApJ...721.1014M}
}

@BOOK{gpr2006,
       author = {{Rasmussen}, Carl Edward and {Williams}, Christopher K.~I.},
        title = "{Gaussian Processes for Machine Learning}",
         year = 2006,
       adsurl = {https://ui.adsabs.harvard.edu/abs/2006gpml.book.....R}
}

@ARTICLE{proga2000,
       author = {{Proga}, Daniel},
        title = "{Winds from Accretion Disks Driven by Radiation and Magnetocentrifugal Force}",
      journal = {\apj},
         year = 2000,
        month = aug,
       volume = {538},
       number = {2},
        pages = {684-690},
          doi = {10.1086/309154},
archivePrefix = {arXiv},
       eprint = {astro-ph/0002441},
 primaryClass = {astro-ph},
       adsurl = {https://ui.adsabs.harvard.edu/abs/2000ApJ...538..684P}
}

@ARTICLE{aromal2023,
       author = {{Aromal}, P. and {Srianand}, R. and {Petitjean}, P.},
        title = "{Time variability of ultra fast BAL outflows using SALT: C IV equivalent width analysis}",
      journal = {\mnras},
         year = 2023,
        month = jul,
       volume = {522},
       number = {4},
        pages = {6374-6393},
          doi = {10.1093/mnras/stad1370},
archivePrefix = {arXiv},
       eprint = {2305.02352},
 primaryClass = {astro-ph.GA},
       adsurl = {https://ui.adsabs.harvard.edu/abs/2023MNRAS.522.6374A}
}

@ARTICLE{hewett2003,
       author = {{Hewett}, Paul C. and {Foltz}, Craig B.},
        title = "{The Frequency and Radio Properties of Broad Absorption Line Quasars}",
      journal = {\aj},
         year = 2003,
        month = apr,
       volume = {125},
       number = {4},
        pages = {1784-1794},
          doi = {10.1086/368392},
archivePrefix = {arXiv},
       eprint = {astro-ph/0301191},
 primaryClass = {astro-ph},
       adsurl = {https://ui.adsabs.harvard.edu/abs/2003AJ....125.1784H}
}

@ARTICLE{Tombesi2010,
       author = {{Tombesi}, F. and {Cappi}, M. and {Reeves}, J.~N. and
         {Palumbo}, G.~G.~C. and {Yaqoob}, T. and {Braito}, V. and {Dadina}, M.},
        title = "{Evidence for ultra-fast outflows in radio-quiet AGNs. I. Detection and statistical incidence of Fe K-shell absorption lines}",
      journal = {\aap},
         year = 2010,
        month = oct,
       volume = {521},
          eid = {A57},
        pages = {A57},
          doi = {10.1051/0004-6361/200913440},
archivePrefix = {arXiv},
       eprint = {1006.2858},
 primaryClass = {astro-ph.HE},
       adsurl = {https://ui.adsabs.harvard.edu/abs/2010A&A...521A..57T}
}

@ARTICLE{cicone2014,
       author = {{Cicone}, C. and {Maiolino}, R. and {Sturm}, E. and {Graci{\'a}-Carpio}, J. and {Feruglio}, C. and {Neri}, R. and {Aalto}, S. and {Davies}, R. and {Fiore}, F. and {Fischer}, J. and {Garc{\'\i}a-Burillo}, S. and {Gonz{\'a}lez-Alfonso}, E. and {Hailey-Dunsheath}, S. and {Piconcelli}, E. and {Veilleux}, S.},
        title = "{Massive molecular outflows and evidence for AGN feedback from CO observations}",
      journal = {\aap},
         year = 2014,
        month = feb,
       volume = {562},
          eid = {A21},
        pages = {A21},
          doi = {10.1051/0004-6361/201322464},
archivePrefix = {arXiv},
       eprint = {1311.2595},
 primaryClass = {astro-ph.CO},
       adsurl = {https://ui.adsabs.harvard.edu/abs/2014A&A...562A..21C}
}

@ARTICLE{green2023,
       author = {{Green}, Kaylie S. and {Gallagher}, Sarah C. and {Leighly}, Karen M. and {Choi}, Hyunseop and {Grupe}, Dirk and {Terndrup}, Donald M. and {Richards}, Gordon T. and {Komossa}, S.},
        title = "{Investigating the Origin of the Absorption-line Variability in the Narrow-line Seyfert 1 Galaxy WPVS 007}",
      journal = {\apj},
         year = 2023,
        month = aug,
       volume = {953},
       number = {2},
          eid = {186},
        pages = {186},
          doi = {10.3847/1538-4357/ace2c4},
       adsurl = {https://ui.adsabs.harvard.edu/abs/2023ApJ...953..186G}
}
\bibliographystyle{aasjournalv7}

%% This command is needed to show the entire author+affiliation list when
%% the collaboration and author truncation commands are used.  It has to
%% go at the end of the manuscript.
%\allauthors

%% Include this line if you are using the \added, \replaced, \deleted
%% commands to see a summary list of all changes at the end of the article.
%\listofchanges

\end{document}